\documentclass[prd,twocolumn,superscriptaddress,floatfix,amsmath,amssymb,nofootinbib,nobibnotes]{revtex4-1}
\usepackage[english]{babel}
\usepackage{graphicx}
\usepackage{dcolumn}
\usepackage{bm}
\usepackage{verbatim}
\usepackage{mathrsfs}
\usepackage{cancel}
\usepackage{epstopdf}
\usepackage{color}
\usepackage[usenames,dvipsnames]{pstricks}
\usepackage{epsfig}
\usepackage{pst-grad} 
\usepackage{pst-plot} 
\usepackage{xcolor}

\usepackage{lipsum}
\usepackage{subfigure}
\usepackage[normalem]{ulem}
\usepackage{hyperref}

\newcommand{\nn}{\nonumber}
\newcommand{\be}{\begin{equation}}
\newcommand{\ee}{\end{equation}}
\newcommand{\bea}{\begin{eqnarray}}
\newcommand{\eea}{\end{eqnarray}}
\newcommand{\e}{\mathrm{e}}\newcommand{\ii}{\mathrm{i}}

\makeatletter
\newcommand{\vast}{\bBigg@{4}}
\newcommand{\Vast}{\bBigg@{5}}
\makeatother

\begin{document}

\title{Probing geometric information using the Unruh effect in the vacuum}
\author{Aida Ahmadzadegan}
\affiliation{Department of Applied Mathematics, University of Waterloo, Waterloo, Ontario, N2L 3G1, Canada}
\affiliation{Perimeter Institute for Theoretical Physics, Waterloo, Ontario N2L 2Y5, Canada}
\author{Fatemeh Lalegani}
\email{Work done while this visitor was at the physics department of the University of Waterloo.}
\affiliation{Department of Physics, Isfahan University of Technology, Isfahan 84156-83111, Iran}
\author{Achim Kempf}
\affiliation{Department of Applied Mathematics, University of Waterloo, Waterloo, Ontario, N2L 3G1, Canada}
\affiliation{Perimeter Institute for Theoretical Physics, Waterloo, Ontario N2L 2Y5, Canada}
\affiliation{Institute for Quantum Computing, University of Waterloo, Waterloo, Ontario, N2L 3G1, Canada}
\affiliation{Department of Physics and Astronomy, University of Waterloo, Waterloo, Ontario N2L 3G1, Canada}
\author{Robert B. Mann}
\affiliation{Department of Physics and Astronomy, University of Waterloo, Waterloo, Ontario N2L 3G1, Canada}
\affiliation{Perimeter Institute for Theoretical Physics, Waterloo, Ontario N2L 2Y5, Canada}

\begin{abstract}
We present a new method by which, in principle, it is possible to ``see in absolute darkness'', i.e., without exchanging any real quanta through quantum fields. This is possible because objects modify the mode structure of the vacuum in their vicinity. The new method probes the mode structure of the vacuum through the Unruh effect, i.e., by recording the excitation rates of quantum systems that are accelerated.  
\end{abstract}

\maketitle
\setlength{\abovedisplayskip}{4pt}
\setlength{\belowdisplayskip}{4pt}

Any quantum system that can act as a detector of field quanta must couple to the field, i.e., it must contain a charge. The Unruh effect then arises  because, as the detector is accelerated, so is its charge and this will generally excite the quantum field. Crucially, at the same time, through the same interaction Hamiltonian, the quantum field can then also excite the detector. For example, a uniformly 
accelerated detector coupled to a quantum field in its Minkowski vacuum will get excited in this way as if exposed to a thermal bath of temperature $T= {\alpha}/2\pi$, where ${\alpha}$ is the magnitude of the detector's proper acceleration \cite{Unruh1976,Sewell1982,Crispino,Birrell1984,Wald1994,Unruh-Wald}. The Unruh effect has been predicted and derived in a broad variety of contexts, and it has been extended  to fields confined within cavities  \cite{cavityquantum2004,Brown2012} and  to  non-uniformly accelerated trajectories \cite{Ostapchuk:2011ud,Doukas2013,Aida22014}. In particular, it has been shown in \cite{Ahmadzadegan:2018bqz} that the Unruh effect is highly sensitive to non-uniformity of the acceleration. 

Here, we explore the possibility that the sensitivity of the Unruh effect can be further exploited, namely to `see' neutral objects in complete darkness, i.e., without the use of real photons. Seeing in complete darkness should be possible because objects influence the structure of the vacuum around them by effectively setting boundary conditions on field modes or, more generally, by leading to a dressing of the vacuum around the objects though virtual photons. The dressing is known to arise because the ground state of a composite system consisting of a localized system of first quantized matter and a quantum field is generally not the tensor product of their respective ground states, due to the presence of their interaction Hamiltonian.  
In principle, any method for detecting the dressing could be used to see in the dark, i.e., to see without real photons, for example, by using the Casimir effect, or perhaps by using the dark port of a quantum homodyne detector to register modulations of the statistics of vacuum fluctuations. 

Here we show that, in principle, seeing in complete darkness can be realized elegantly by using just a single non-uniformly accelerated qubit, i.e., by using the Unruh effect. 

The general detector model we employ is an Unruh-DeWitt detector (UDW) \cite{DeWitt,Birrell1984,Hu:2012jr}, an idealized model of a real particle detector that encompasses all fundamental features of the light-matter interaction when there is no angular momentum exchange involved \cite{Wavepackets}. It consists of a localized two-level quantum system (a qubit) that linearly couples to a scalar field. For examples of studies  of the response of UDW detectors in Minkowski and curved spacetimes, see, e.g.,    \cite{Satz2006,Obadia,Hodgkinsonclick,Jennings:2010vk,Ng:2018drz,Aida12014,ng2014unruh}. 

Our goal is to determine the response of such detectors undergoing various non-uniform acceleration regimes
inside an optical cavity of proper length $L$ with reflecting boundary conditions. As we show, the detector's response can be used to infer the location of the boundary of the cavity. In other words, the presence and structure of the cavity can be inferred without any exchange of real quanta, and so can be seen, in this sense, in complete darkness. For the trajectories of the detector, there are of course many choices. The literature on UDW detectors, apart from the standard uniformly accelerated and asymptotically null cases \cite{Davies}, considers for example trajectories for which a constant energy flux is emitted \cite{Carlitz:1986nh}. There are also several asymptotically inertial trajectories \cite{1982JPhA...15L.477W} that have been considered, and yet other trajectories possess the virtue that their cases are exactly solvable and exhibit interesting physical features \cite{Goodthesis}.  

\section{Settings}

\begin{figure}[htp]
\includegraphics[scale=0.45]{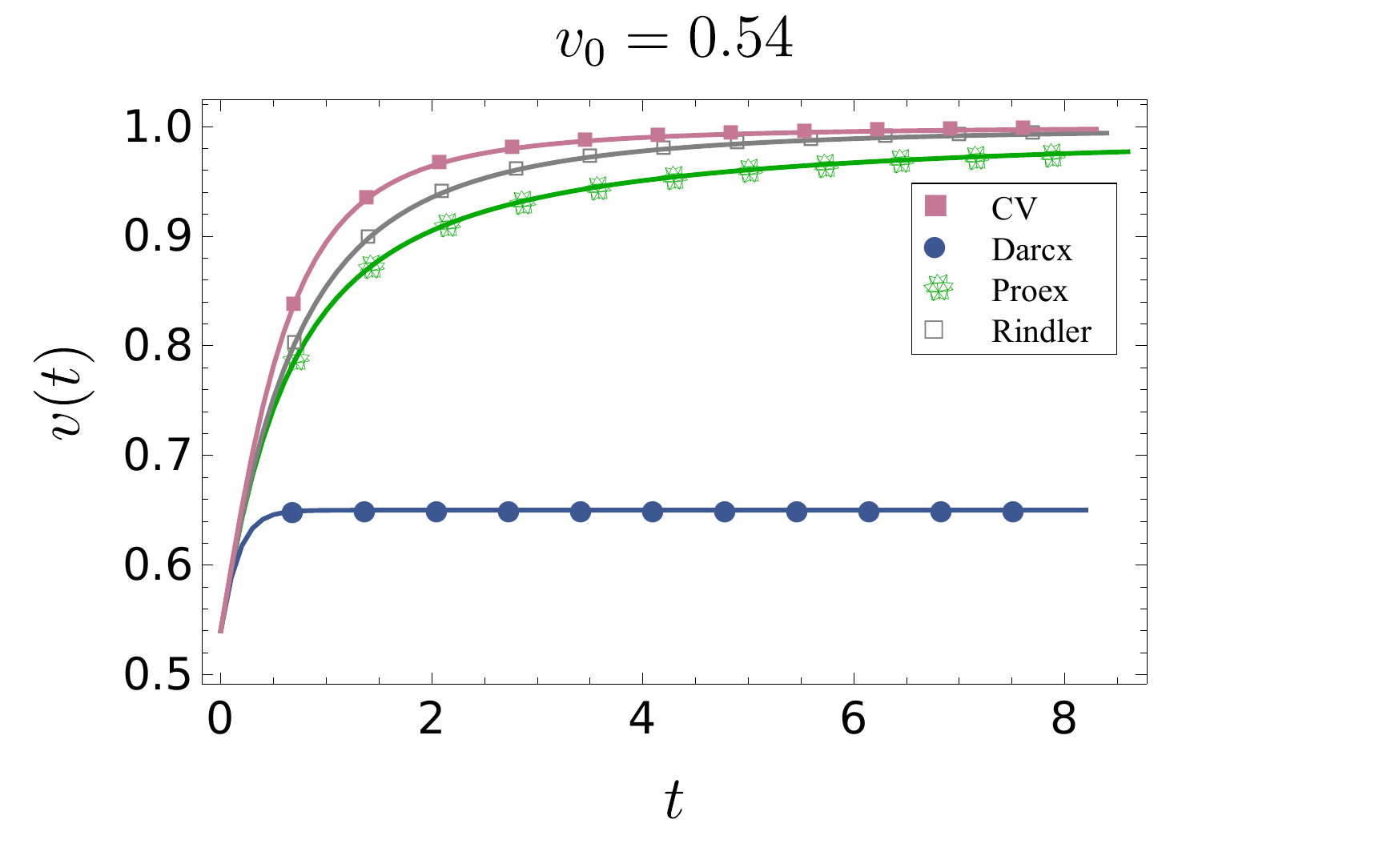}
\includegraphics[scale=0.45]{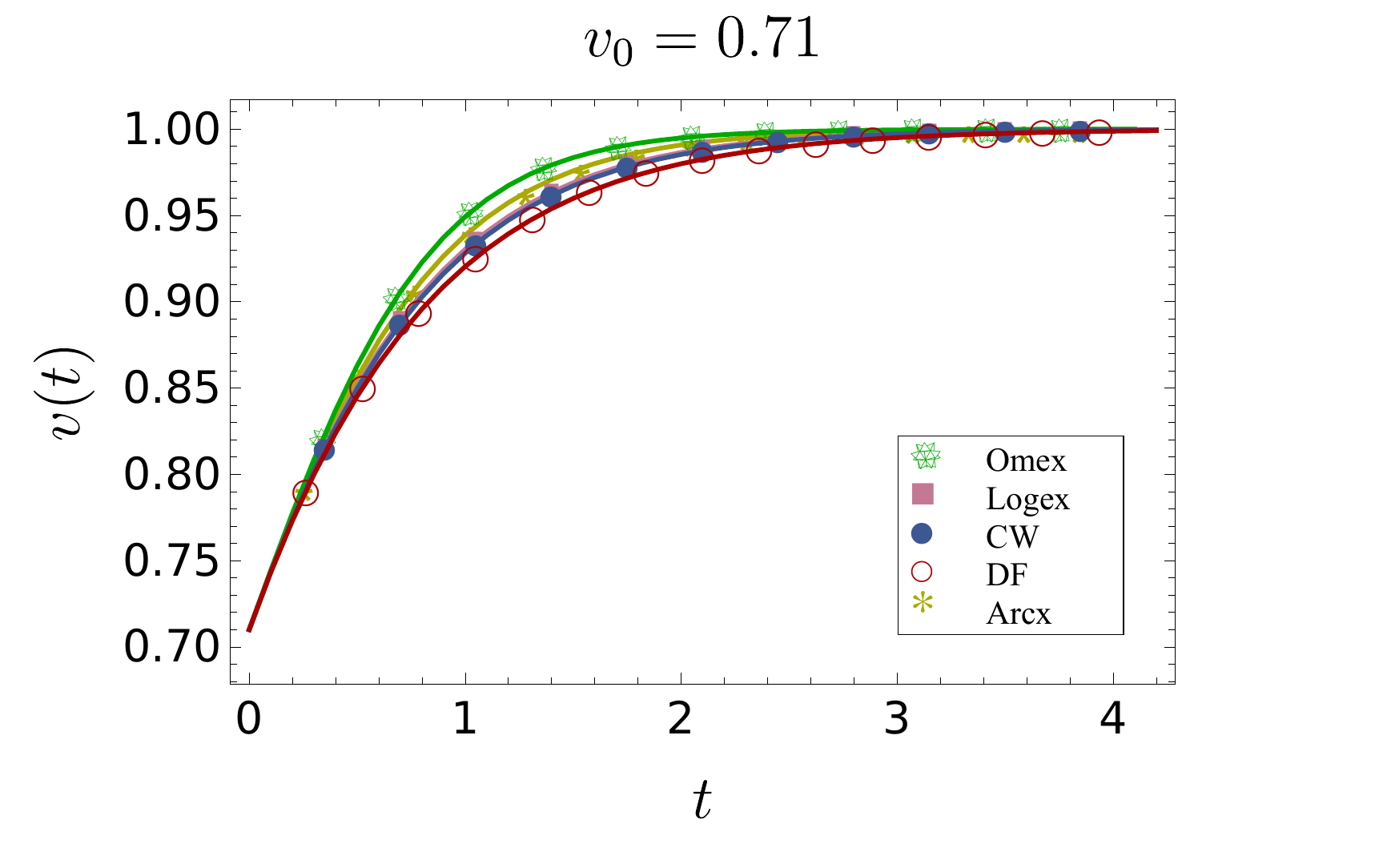}
\caption{Velocities for the various trajectories of a) vanishing flux: Rinder, Costa-Villalba (CV), Darcx, and Proex and b) finite flux: Davies-Fulling (DF), Arcx, Omex, Logex, and Carlitz-Willey (CW). Initial velocities of all trajectories in (a) and (b) are chosen to be $v_0=0.54$ and $v_0=0.71$, respectively so that all trajectories and their parameters stay positive and physical. The acceleration in the Rindler case has been normalized to $\alpha_0=1$. Initial accelerations of the other trajectories are normalized to this value.}
\label{vtraj}
\end{figure}

We shall adopt these various trajectories to address the problem of interest, analyzing the excitation probability of a UDW detector along a given non-uniformly accelerating trajectory whilst inside a cavity. Though we work in (1+1) dimensions, our results can be straightforwardly extended to higher dimensions. We compare the result of each case with the excitation probability of a detector moving on the Rindler trajectory. We further classify the motions into two broad categories: trajectories with vanishing asymptotic flux and trajectories with a finite asymptotic flux; we depict the respective velocity profiles of  these trajectories in Figs.~\ref{vtraj}(a) and \ref{vtraj}(b).
For each case we analyze the motion of the detector from the cavity frame as it travels  through the cavity. Throughout, Minkowski coordinates  in  the cavity's rest frame are denoted as $(x,t)$, and  $\tau$ denotes the proper time of the detector; we follow the convention of setting $c=\hbar=1$. For any trajectory $x=x(t)$, it is straightforward to define the following quantities associated with the motion of the detector,

$$
\begin{array}{lll}
\!v &\!\!\!= \frac{\mathrm{d}x}{\mathrm{d}t}  &\textrm{coordinate velocity}  \\
\!\tau &\!\!\!= \int  \sqrt{1-v^2} \mathrm{d}t &\textrm{detector proper time}  \\
\!u^\mu &\!\!= \left(\frac{1}{\sqrt{1-v^2}} ,\frac{v}{\sqrt{1-v^2}}\right) &\textrm{detector 2-velocity}  \\
\!a^\mu &\!\!=  \frac{\mathrm{d}u^\mu}{\mathrm{d}\tau} =\! \left(\frac{v}{(1-v^2)^2},\frac{1}{(1-v^2)^2}\right)  \!\frac{\mathrm{d}v}{\mathrm{d}t}&\!\textrm{detector 2-acceleration}  \\
\!\alpha &\!\!\!= \sqrt{a^\mu a_\mu} = \frac{1}{(\sqrt{1-v^2})^3}\frac{\mathrm{d}v}{\mathrm{d}t} &\textrm{proper acceleration}  
\end{array}
$$

We are particularly interested in comparing detector responses between various trajectories, each of which has the detector entering the optical cavity at  $\big(t_0,x(t_0) \big)$. Therefore, we must calibrate the motions of the detectors as they enter the cavity so that they all begin with the same initial velocity (so as to remove spurious Doppler effects)  and the same initial acceleration (so as to properly compare to the uniformly accelerated case).  Imposing these constraints fixes the initial time parameter $t_0$ and the acceleration parameter for each of the motions under consideration. For each trajectory, the ratio  of its acceleration $\alpha(t)$ relative to the uniform Rindler case is a monotonically increasing function of $t$. Further description of all the trajectories is given in the appendix.

Our comparisons of the detector responses are within each category of trajectories, since each produces qualitatively distinct responses at late times. For each trajectory, we measure the excitation probability $P$ of the detector for a period of time $T$ as it traverses the full proper length $L$ of the optical cavity. They are all calibrated to have the same initial velocities and accelerations as the Rindler case. The time evolution of the system follows the atom-field Hamiltonian that generates evolution for the entire system with respect to the time coordinate $t$ of the cavity's proper frame \cite{Brown2012} given by

\bea\label{Hamiltonian2}
\hat{H}(t)=\frac{\mathrm{d}\tau}{\mathrm{d}t}\hat{H}^{(\text{d})}_{\text{free}}[\tau(t)]+\hat{H}^{(\text{f})}_{\text{free}}(t)+\frac{\mathrm{d}\tau}{\mathrm{d}t}\hat{H}_{\text{int}}[\tau(t)],  
\eea
where $\hat{H}_{\text{I}}[\tau(t)]= \lambda \chi(t) \hat{\mu}(t) \hat{\phi}\left[x(t)\right]$ models the detector-field interaction \cite{Unruh1976,DeWitt,Crispino,Louko2008,Satz2006}.  The constant $\lambda$ is the coupling strength, $\chi(t)\ge0$ is the window function, switching the interaction on and off,  $\hat{\mu}(t)$ is the monopole moment of the detector, and $\hat{\phi}\left[x(t)\right]$ is the massless scalar field that the detector is interacting with along its trajectory. The monopole moment operator takes the usual form of $\hat{\mu}(t)=\big(\sigma^{+}\e^{\ii\Omega\tau(t)}+\sigma^{-}\e^{-\ii\Omega \tau(t)}\big)$, in which $\Omega$ is the proper energy gap between the ground state, $\left| g \right\rangle$ and the excited state, $\left| e \right\rangle$ of the detector and $\sigma^{\pm}$  are ladder operators ($\sigma^{+}\left| g \right\rangle = \left| e \right\rangle$, $\sigma^{-}\left| e \right\rangle = \left| g \right\rangle$). Working in this frame and expanding the field in terms of an orthonormal set of solutions to the Schr\"{o}dinger equation inside the cavity yields the following Hamiltonian 

\be \label{hamilto2}
\hat{H}_{\text{I}}(t)=\lambda \frac{\mathrm{d}\tau}{\mathrm{d}t}\sum^{\infty}_{n=1}\frac{\hat{\mu}(t)}{\sqrt{\omega_n L}}\big(\hat{a}^{\dagger}_{n}u_n [x(t),t]+\hat{a}_{n}u^{*}_n [x(t),t]\big)
\ee

in the interaction picture. We consider Dirichlet (reflective) boundary conditions  $\phi\left[0,t\right]=\phi \left[L,t\right]=0$, and so the field modes take the form of the stationary waves $u_n [x(t),t]=\e^{i\omega_n t}\sin [k_n x(t)]$. Here, $\omega_n^2=k_n^2+m^2$ where $k_n=n \pi/L$. In our study, we work with scalar fields, therefore, $m=0$.

To characterize the vacuum response of a particle detector undergoing different trajectories, we initially prepare the detector in its ground state and the cavity in the vacuum state, so
that its initial density matrix is 
$\rho_0=\left|g\right\rangle\left\langle g\right|\otimes \left|0\right\rangle\left\langle 0\right|$. The time evolution of the system is governed by the interaction Hamiltonian \eqref{hamilto2} in the time interval $0<t<T$ and is given by $\hat{U}\equiv \hat{U}(T,0)=\mathcal{T} \e^{-\ii \int{\mathrm{d}\tau \hat{H}_{I}(\tau)}}$. We consider the coupling constant, $\lambda$, to be a small parameter\footnote{ Note that in $(1 + 1)$ dimensions, the coupling constant has units of inverse length in natural units. Here, small coupling strength means the dimensionless quantity, $\lambda \sigma$ is small, where $\sigma=1$ is the fiducial unit length of the cavity; all length scales are in units of $\sigma$.} so we can work within the validity of perturbation theory. Therefore, using the Dyson perturbative expansion up to second order in $\lambda$, we can write \cite{AasenPRL}, 

\be
\rho_{T}\!=\!\big[\openone+\hat{U}^{(1)}\!+\hat{U}^{(2)}\!+\mathcal{O}(\lambda^3)\big]\rho_0\big[\openone+\hat{U}^{(1)}\!+\hat{U}^{(2)}\!+\mathcal{O}(\lambda^3)\big]^{\dagger}
\ee
where

\begin{align}
&\hat{U}^{(1)}\!\!=\!\frac{\lambda}{i}\!\sum^{\infty}_{n=1}\!\!\big[\sigma^{+} a^{\dagger}_{n}I_{+,n}\!+\!\sigma^{-} a_{n}I^{\ast}_{+,n}
\!+\!\sigma^{-} a^{\dagger}_{n}I_{-,n}\!+\!\sigma^{+}a_{n}I^{\ast}_{-,n}\big] \nonumber\\
&I_{\pm,n}=\int^{T}_0 \!\frac{\mathrm{d}\tau}{\mathrm{d}t}\e^{\ii\left[\pm \Omega \tau(t)+\omega_nt\right]} \sin\left[k_n \big(x(t)-x(t_0)\big)\right]dt. 
\end{align}

We compute the density matrix  $\rho_{T, (\text{d})}$ for the detector by taking the partial trace over the field degrees of freedom  \cite{AasenPRL}. The first order contribution to the  transition probability vanishes, so the leading contribution comes from second order in the coupling strength. Therefore, the excitation probability of the detector is
\bea\label{P}
P=\lambda^2\sum_{n=1}^{\infty}\left|I_{+,n}\right|^2
\eea
We work with this quantity
rather than  the transition rate as there is no formal or computational advantage in the latter given the absence of time translation invariance in our setting;  both quantities contain the same information.

\section{Results}

\begin{figure}[htp]
\includegraphics[width=0.45\textwidth]{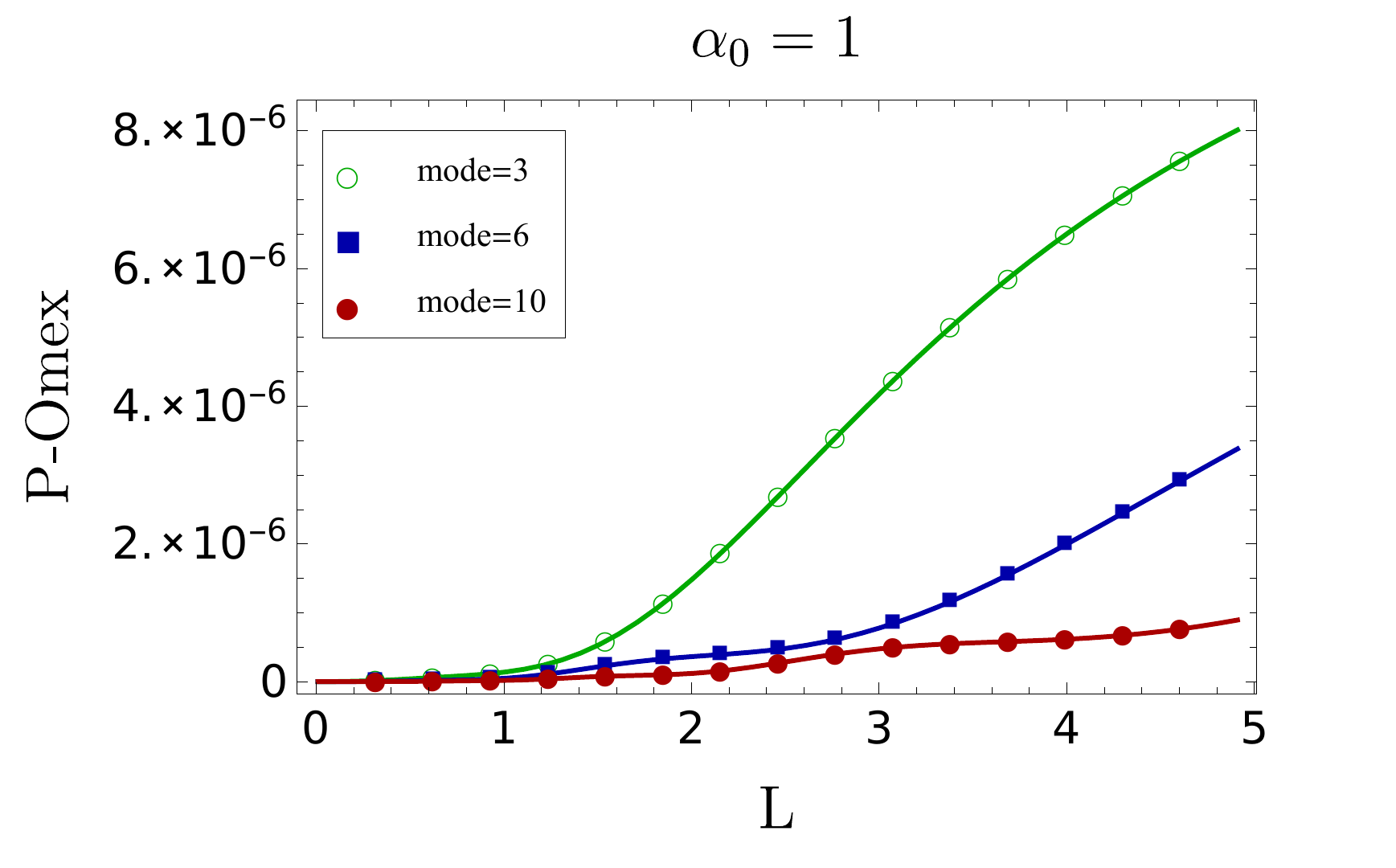}	
\caption{Plots of the excitation probability of the detector moving on the Omex trajectory with initial acceleration ratio $\alpha_0=1$. The detector gap is chosen such that it resonates with the third mode of the field (empty circle), with the sixth mode of the field (square), and with the tenth mode of the field (full circle).}
\label{omexmodes}
	\end{figure}

In obtaining our results, there are a few factors that determine the response of the detector inside the cavity. We keep the coupling constant small ($\lambda=0.01$) and choose the gap of the detector to be in resonance with one of the field modes inside the cavity. By changing the resonance mode (choosing a different gap for the cavity) the behaviour of the excitation probability $P$ changes. For example, in Fig.~\ref{omexmodes},  $P$ for a  detector moving on the Omex trajectory is given as a function of the cavity length for three different values of resonance mode. As we can see, the excitation probability of a detector in resonance with lower modes of the field shows more sensitivity to the change in length of the optical cavity, and so is a preferred choice for inferring the location of its boundary.
	
\begin{figure}[htbp]
\centering
\includegraphics[scale=0.45]{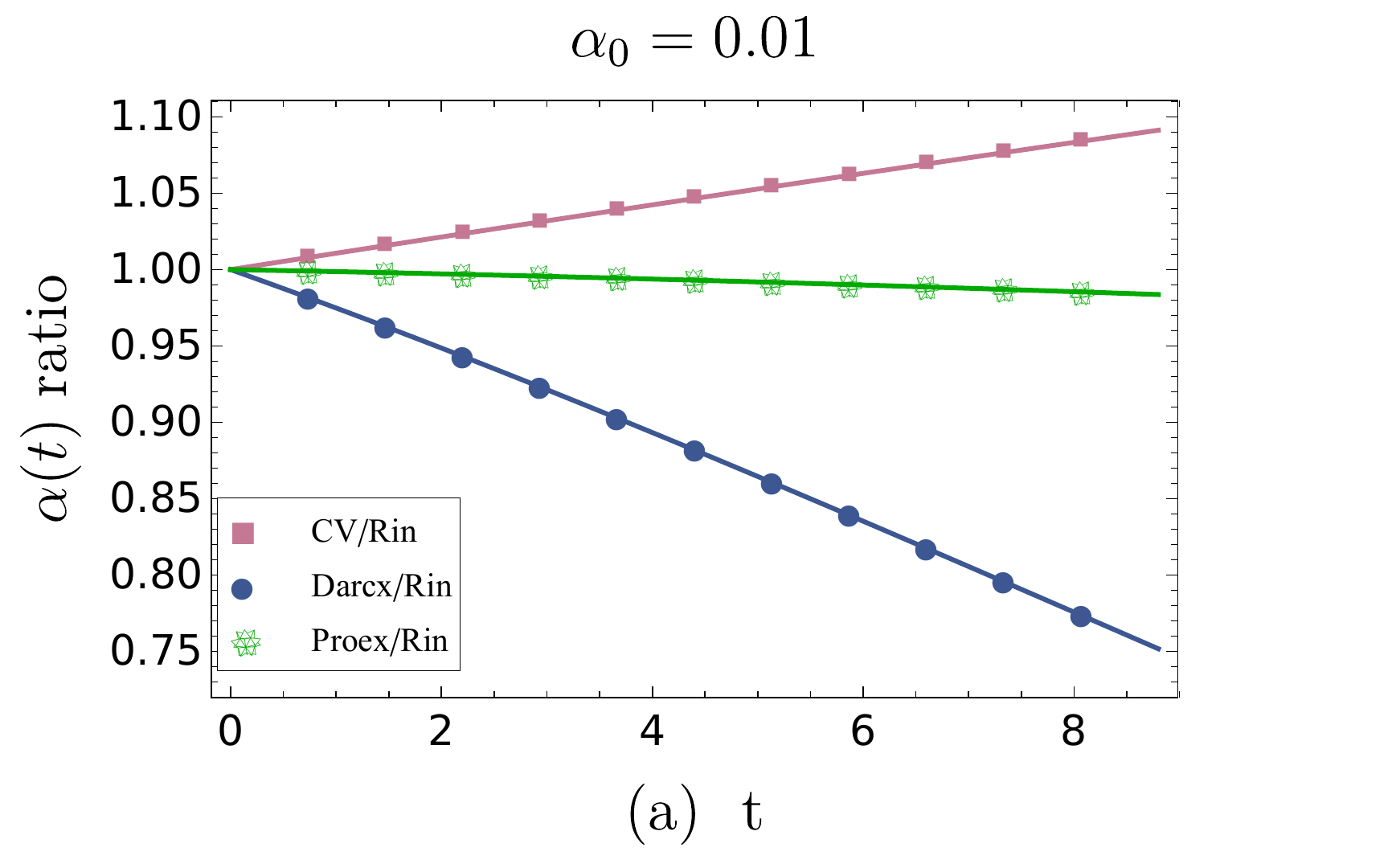}
\includegraphics[scale=0.45]{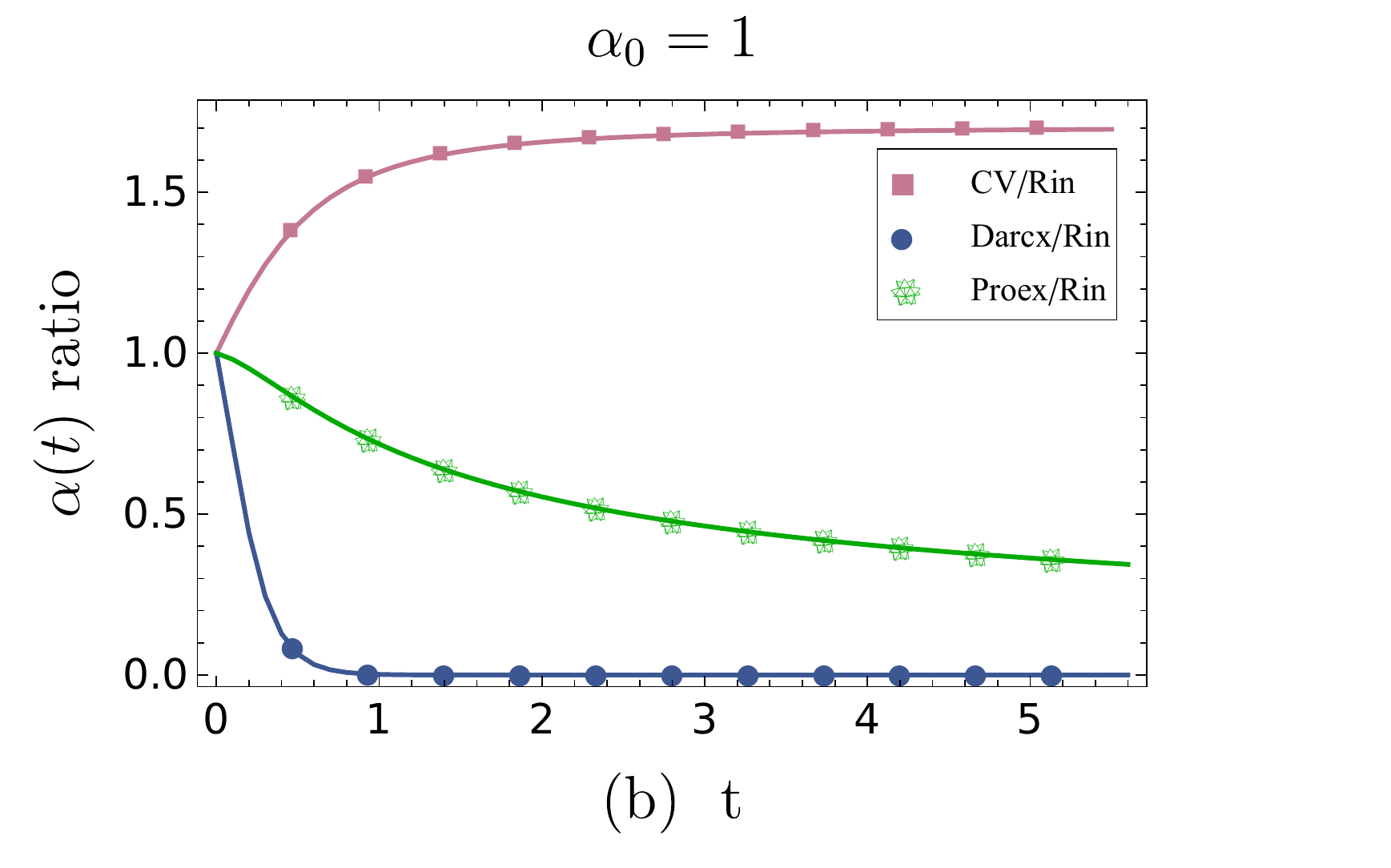}
\caption{The three acceleration ratios are CV (pink square), Darcx (blue circle), and Proex (green star) to Rindler. Each plot depicts the behaviour of these ratios as a function of the cavity length for initial accelerations: a) $\alpha_0=0.01$ and b) $\alpha_0=1$.}
\label{vanisha}
\end{figure} 

\begin{figure}[htbp]
\centering
\includegraphics[scale=0.45]{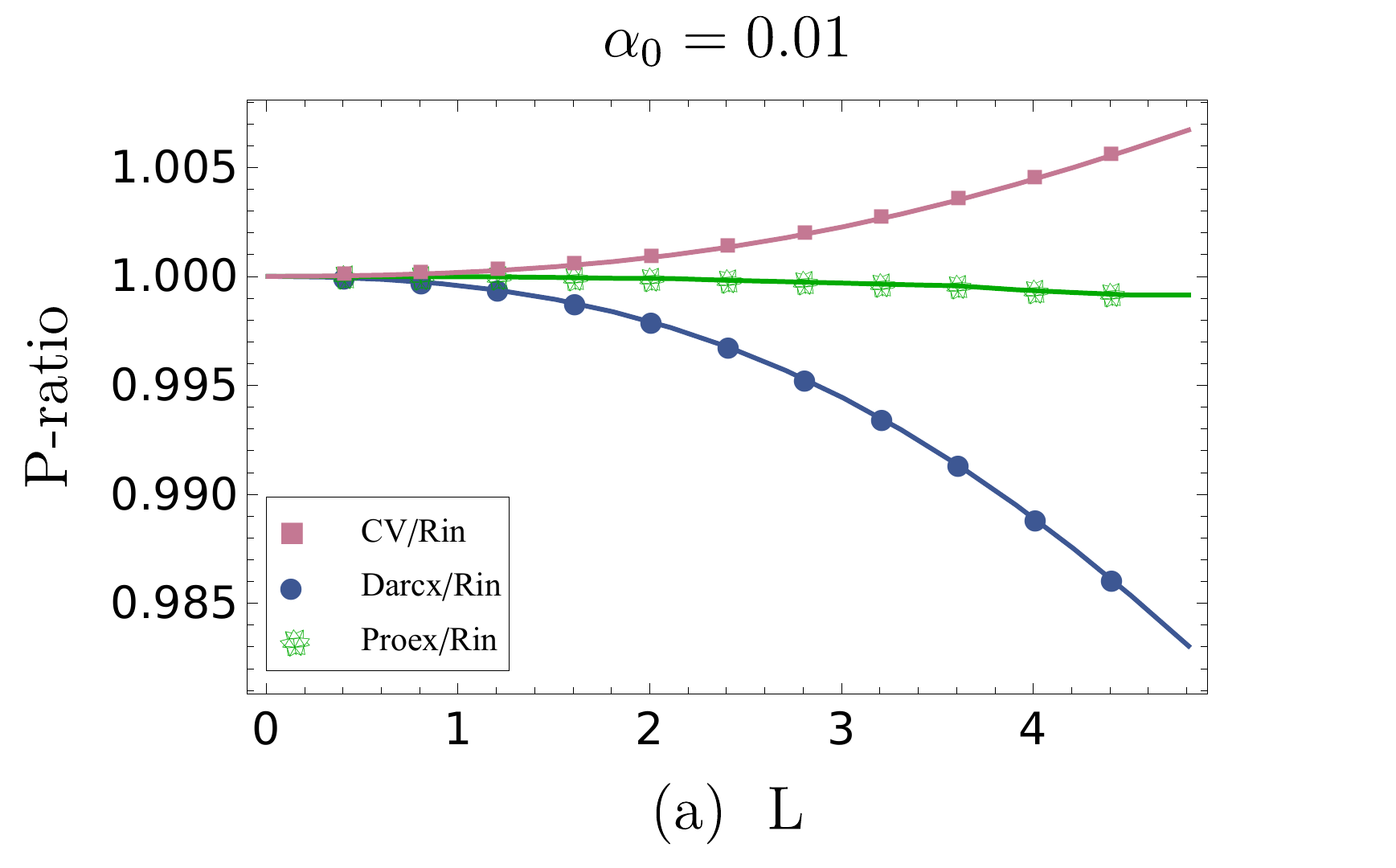}
\includegraphics[scale=0.45]{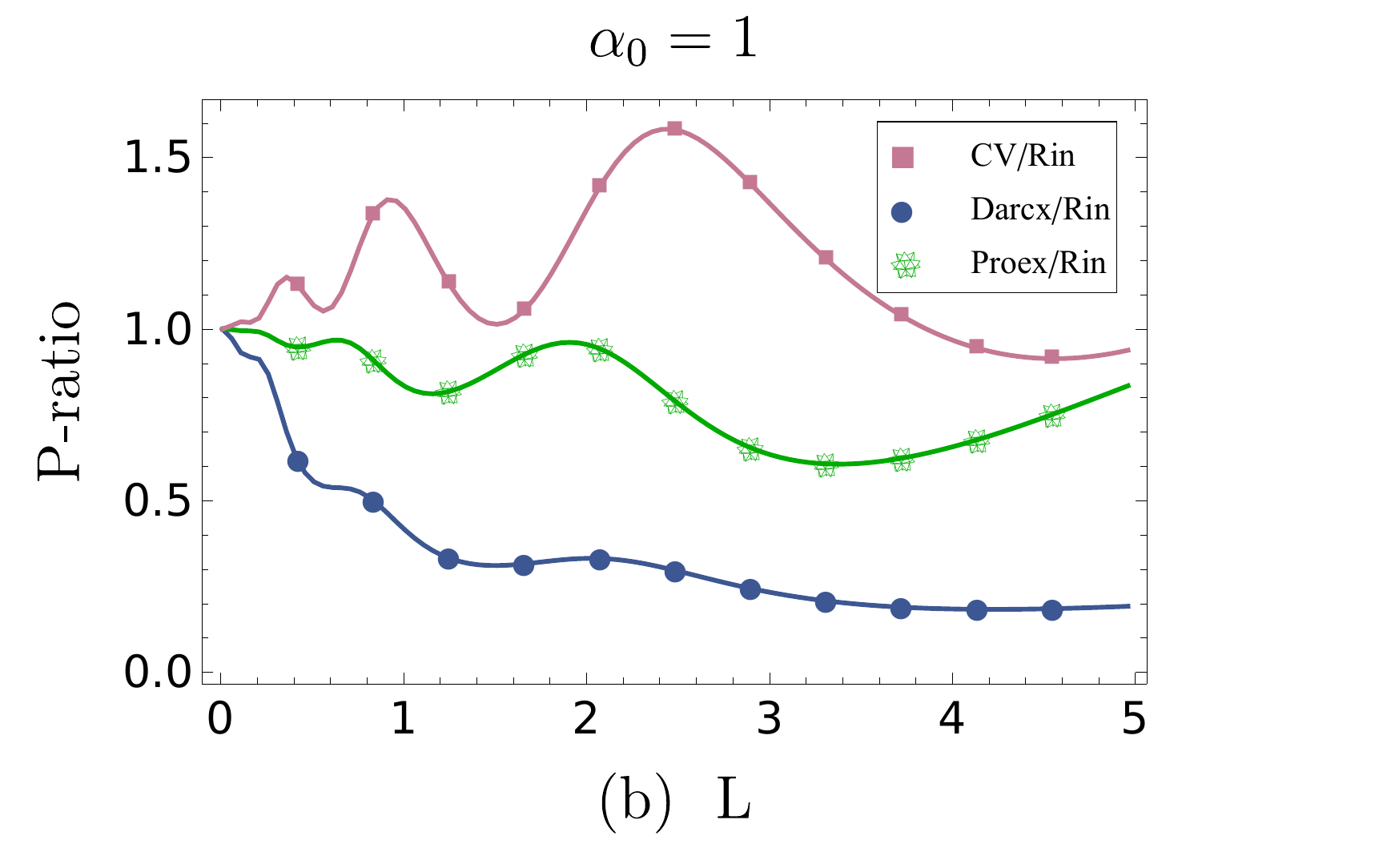}
\caption{The three transition probability ratios are CV (pink square), Darcx (blue circle), and Proex (green star) to Rindler. Each plot represents the behaviour of these ratios as a function of the cavity length that they are traveling through with different initial accelerations: a) $\alpha_0=0.01$ and b) $\alpha_0=1$.}
\label{vanishp}
\end{figure}

\begin{figure}[htbp]
\centering
\includegraphics[scale=0.45]{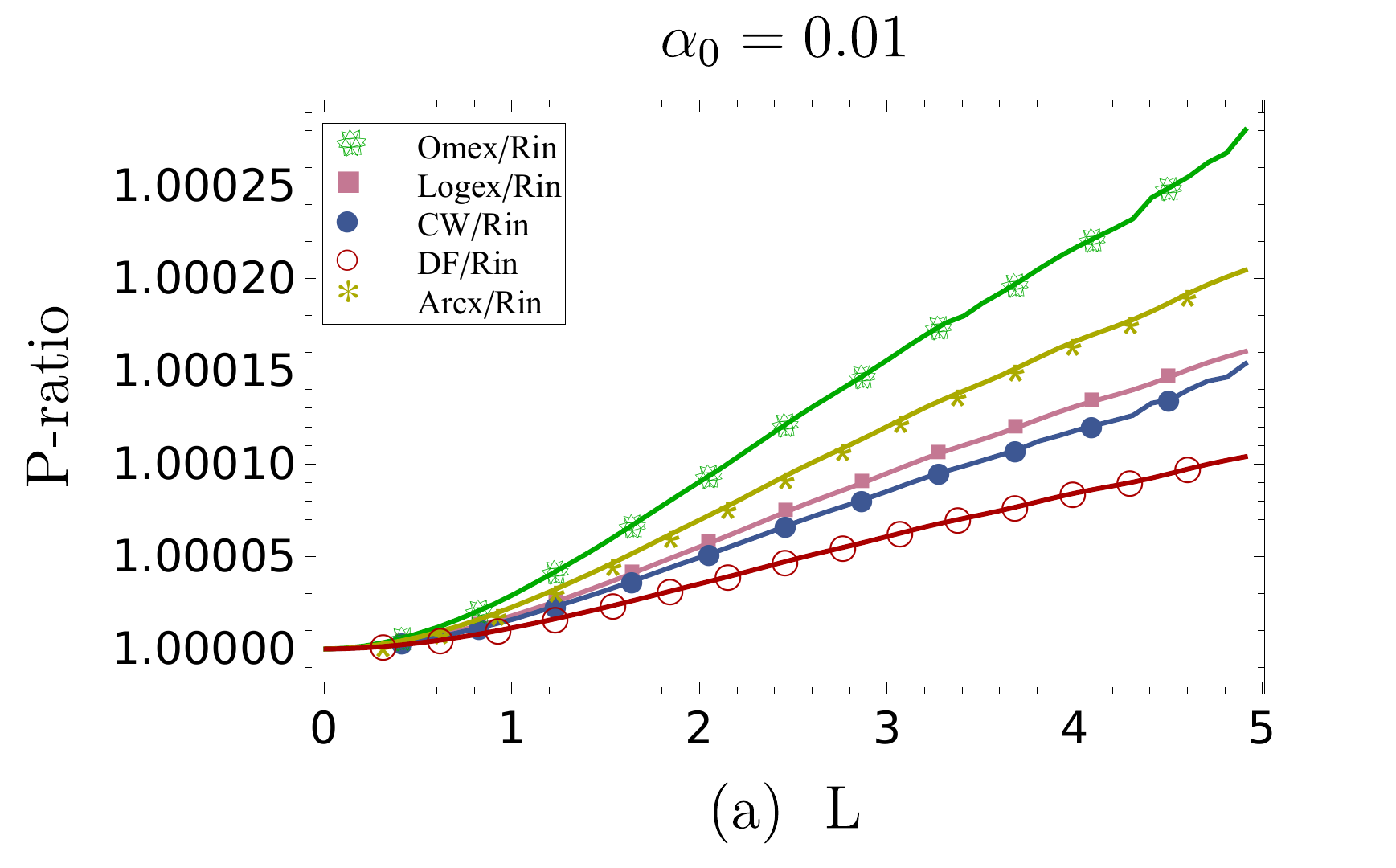}
\includegraphics[scale=0.45]{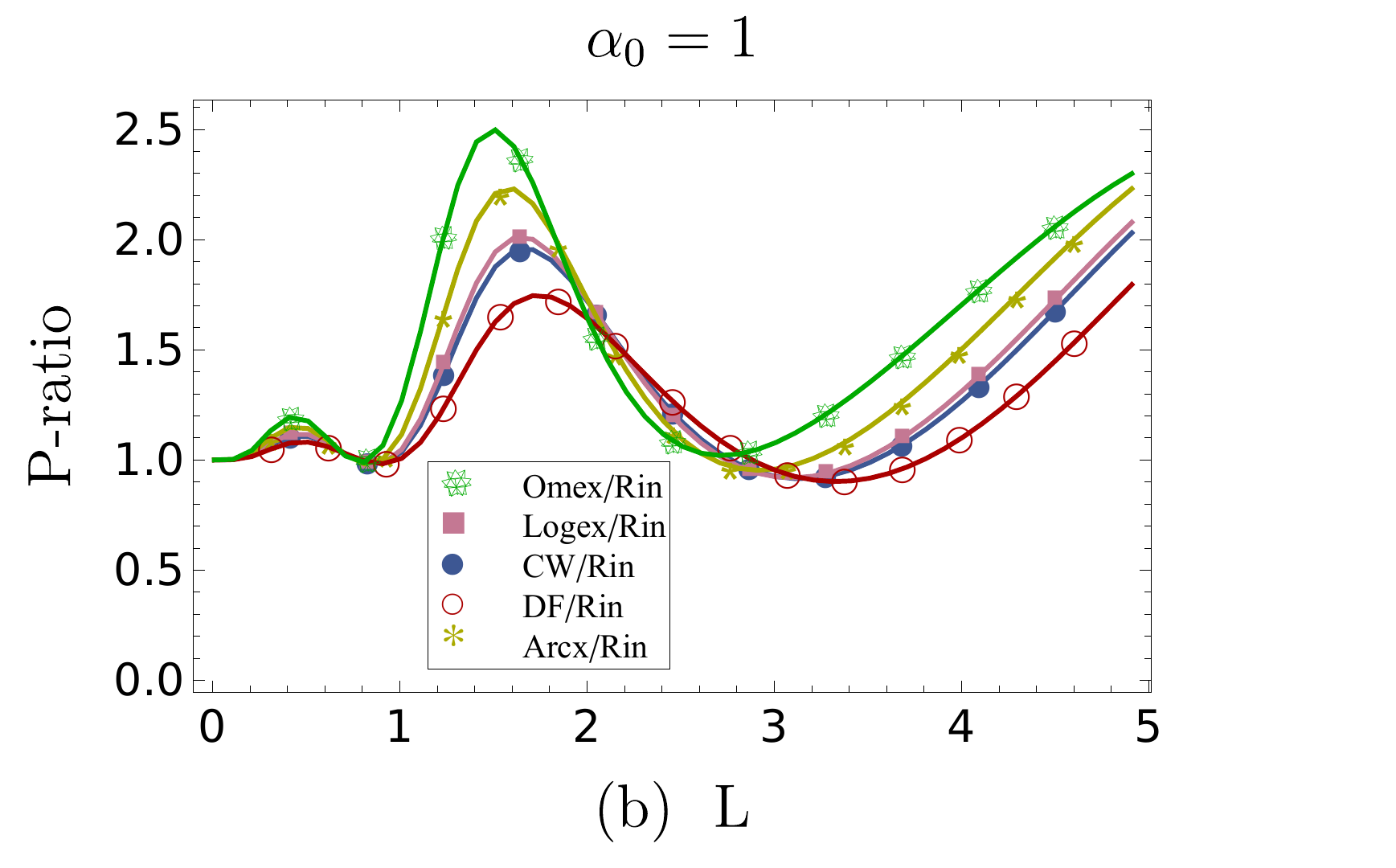}
\caption{The five transition probability ratios are Omex (in green flake), Logex (in pink square), CW (in blue circle), DF (in red circle), and Arcx (in yellow star) to Rindler. Each plot represents the behaviour of these ratios as a function of the cavity length that they are traveling through with different initial accelerations: a) $\alpha_0=0.01$ and b)$\alpha_0=1$.}
\label{nonvanishp}
\end{figure}

In the following plots we depict the ratio of excitation probabilities of a UDW detector moving on different non-uniform trajectories relative to the uniformly accelerated Rindler case.  We find that varying choices of accelerated trajectories are differently suited for the task of seeing in the dark.
 In Fig.~\ref{vanishp} we plot as a function of the cavity's proper length $L$ (in units of inverse gap frequency) the excitation probability ratio (P-ratio) of a detector moving on CV, Darcx, and Proex trajectories relative to the excitation probability of a detector moving on a Rindler trajectory, with initial velocity $v=0.54$ and two different initial proper accelerations $\alpha(t=t_0) \equiv \alpha_0$  over the range $0.01<L<5$. Similarly, Fig.~\ref{nonvanishp} presents the P-ratio of a detector moving on Omex, Logex, CW, DF, and Arcx trajectories relative to the Rindler trajectory, with initial velocity $v=0.71$. For each trajectory we calibrate the detector so that it enters the cavity at $t_0$ with the same initial acceleration and velocity as that of the Rindler trajectory; for each case the gap of the detector is in resonance with the sixth mode of the field. For small initial acceleration, there is a correlation between the acceleration ratio as shown in Fig.~\ref{vanisha}(a) and detector response ratio illustrated in Fig.~\ref{vanishp}(a).

However, as the initial acceleration increases, interesting structure emerges in the detector response ratio as a function of cavity length, indicated in  Fig.~\ref{vanishp}(b). The monotonic behaviour of  $\alpha(t)$ in Fig.~\ref{vanisha}(b) does not yield monotonicity of the detector response ratio -- indeed, we see that it oscillates, decreasing over certain ranges of $L$ despite the increase in  
$\alpha(t)$. This behaviour is the result of choosing the energy gap of the detector to resonate with a specific field mode for all trajectories. 
Similar behaviour in the finite flux case is illustrated in Figs.~\ref{nonvanishp}(a) and \ref{nonvanishp}(b). More examples with different initial accelerations are given in the appendix. In general, a detector 
goes out of resonance with the field mode at a different time (position along the cavity) than for the Rindler trajectory, leading to a distinct signature for a given trajectory with given boundary conditions. Furthermore, one can see the sensitivity to the non-uniformity of acceleration \cite{Ahmadzadegan:2018bqz}, with the Omex trajectory indicating the greatest sensitivity for the mode in question. 
Given a specific non-uniform trajectory, its value of $P$ is sensitive to the length of the cavity and thus  sensitive to the location of each of its boundaries.
 This sensitivity can be exploited to detect the cavity boundaries, without any exchange of real quanta, by only measuring the relative response rate (the P-ratio) of the detector.  
In other words, we can `see in absolute darkness' by only  probing the vacuum field, without sending any signal or radiation.


\section{Outlook}

Our results point towards generalization to sharp vision in all directions in complete darkness, leading to an intriguing close relationship to the field of spectral geometry. One branch of spectral geometry asks, for example, to what extent the  geometry of a Riemannian manifold can be inferred from the spectrum of differential operators on the manifold \cite{Milnor542,kempfspectral}. A related but different branch of spectral geometry asks, for example, to which extent the shape of a drum is encoded in the spectra of the sound it makes \cite{drumspectrum}. 

In our context here, let us 
consider an optical cavity of arbitrary (e.g., convex) shape. This cavity then possesses a corresponding  normal mode decomposition of standing waves of the quantum field, with the shape of the cavity determining the pattern of these standing waves. By sending in multiple detectors  with different energy gaps  moving on varying accelerated trajectories, their
excitation rates will provide information about the cavity boundaries in
various directions, and could therefore  allow the detection, in complete darkness, of the full geometry of the cavity. In this way, an equivalence could be established between the geometry of the cavity and the quantum fluctuations of a quantum system. The establishment of any equivalence between curved shapes or geometries as they occur in general relativity on one hand, and quantum phenomena, such as excitation rates, on the other hand, could ultimately be useful for quantum gravity.

Finally we note that our work may have longer-term applications for short-range sensing.  Indeed, in the absence of a cavity the $P$-ratio of trajectories will be sensitive to the proximity of objects in free space. This is because each object will furnish a boundary condition for the field, or more generally, it will create a dressed quantum vacuum around it. We showed that this change of the dressing of the vacuum can be sensed by accelerated detectors. It will be very interesting to determine the type of trajectories that possess the optimally suited $P$-ratios for such sensing tasks, also in higher dimensions and for massive fields.

\acknowledgements 

A.K. and R. B. M. are supported in part by the Discovery Grant Program of the Natural Sciences and Engineering Research Council of Canada.

\onecolumngrid
\appendix

\section{Vanishing Flux Mirror Trajectories}

\begin{figure}[htp]
\centering
\includegraphics[scale=0.48]{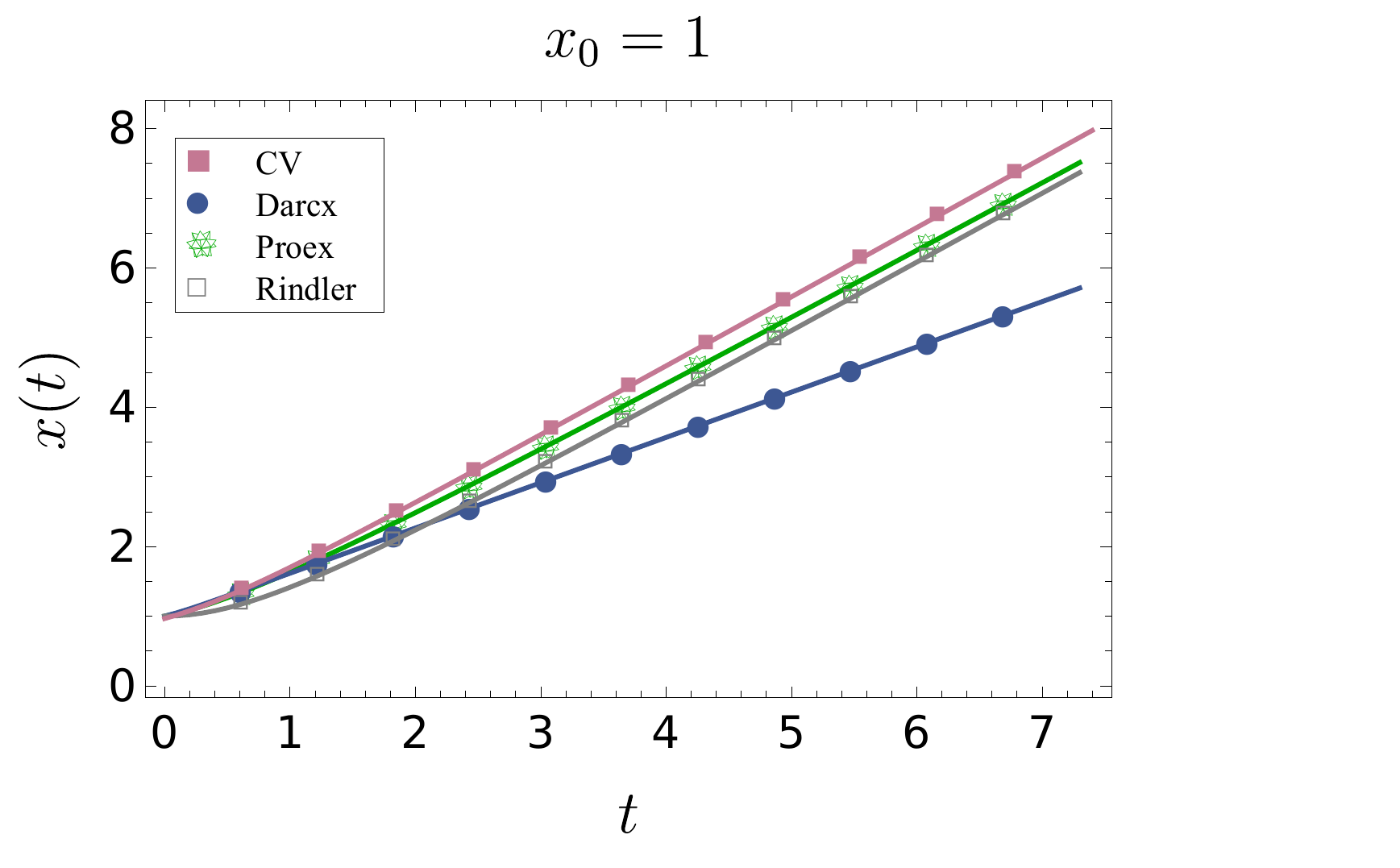}
\caption{Various trajectories of vanishing flux: Rinder, CV, Darcx, and Proex are shown. Initial position of all trajectories is chosen to be $x_0=1$ so that all trajectories and their parameters stay positive and physical. The acceleration in the Rindler case has been normalized to $\alpha_0=1$. Initial accelerations of the other trajectories are normalized to this value.}
\label{traj1-x}
\end{figure}

We begin with a description of the trajectories given in Fig.~\ref{traj1-x} that have vanishing flux.  Their velocities
are presented in the main text in Fig.~1(a). In all the following trajectories, $T$ is measured in the frame of the cavity $(x,t)$.

\subsection{Rindler trajectory}

This is the most commonly studied detector trajectory, 
\be \label{Rintraj}
x(t) = \sqrt{t^{2}+\frac{1}{a^{2}}}, \qquad 
\tau(t) = \frac{1}{a}\text{arcsin}(a t),
\ee
where $\alpha=a$ is the proper acceleration.  It is known that the response rate corresponds to that of a detector in a thermal bath of scalar radiation with temperature $T= a/2\pi$. The detector enters the cavity  with an initial velocity $v_0$. Given its  initial velocity and acceleration,  the time $T$ that it takes for the detector to travel the whole length of the cavity $L$ is
\begin{equation}
 T=\sqrt{\frac{2L\sqrt{1+a^2 t_0^2}+L^2+t_0^2}{a}} 
\end{equation}
obtained by setting  $x(t)=x(t_0)+L$ in \eqref{Rintraj}.

\subsection{Costa-Villalba trajectory}
 
This trajectory is one for which the detector has inertial motion in the distant past and increasingly accelerates to attain uniform acceleration in the distant future along an asymptotically null trajectory \cite{Ostapchuk2008}. The parametrization of the CV trajectory is given by
\bea
x(t)&=&\frac{tw+\sqrt{2+w^2t^2}}{2w},\\\nn
\tau(t)&=&\frac{\text {arcsinh}(tw+\sqrt{2+t^2w^2})}{w}-\frac{\sqrt{1+\frac{1}{(tw+\sqrt{2+t^2 w^2})^2}}}{w},
\eea
where $w$ is a positive constant in CV trajectory whose value is proportional to the uniform acceleration
asymptotically attained at late times. Choosing the same initial acceleration  and  velocity as for the Rindler case, we obtain  
\bea
\!\!\!\!T&=&\frac{1}{2L^2w^2+2Lt_0 w^2-1}\Big(\sqrt{2+t_0^2w^2}(L^2 w+L t_0 w)\\\nn
&+&2L^3w^2+3L^2t_0w^2+Lt_0^2w^2-2L-t_0\Big).
\eea
for the time that it takes for the detector to travel the length $L$ of the cavity. 

The proper acceleration for this trajectory is 
\begin{equation}
\alpha(t)=\frac{w}{\left(1+\frac{1}{(t w+\sqrt{2+t^2 w^2})^2}\right)^{3/2}},
\end{equation} 
and as $t\to \infty$ it is straightforward to show that $\alpha \to w$.

\subsection{Proex}

The Proex trajectory
\bea
x(t)&=&\frac{W(\e^{\sigma t})}{\sigma} \\\nn
\tau(t)&=&\frac{2\sqrt{2 W(\e^{\sigma t})+1} + \ln\left[\frac{\sqrt{2 W(\e^{\sigma t})+1}-1}{\sqrt{2 W(\e^{\sigma t})+1}+1}\right]}{\sigma},
\eea
where $W$ is the product log or Lambert-W function, 
 is a trajectory for which  there is a finite number of particles occupying each mode. Its mirror trajectory has a finite, nonzero energy flux  that vanishes at late times.  
 Both its proper acceleration and acceleration vanish in the distant past and future and in the future the magnitude of the velocity  approaches the speed of light. 
 
 The proper acceleration for this trajectory is given by
\begin{equation}
\alpha(t)=\frac{\sigma W(\e^{\sigma t})}{(2 W(\e^{\sigma t})+1)^{3/2}},
\end{equation} 
and it takes time
  \bea
T=\frac{\log\left( \e^{L\sigma +W(\e^{\sigma t_0}) }(L\sigma +W(\e^{\sigma t_0}))\right) }{\sigma}
\eea
to travel the full length of the cavity.  Both $\sigma$ and $t_0$ are fixed by choosing the initial acceleration and velocity of the Proex detector to be equal to that of the Rindler detector. 

\subsection{Darcx}

The Darcx trajectory, given below, is asymptotically inertial in the past and future, but is not necessarily asymptotically static in the future. A finite amount of particles and energy is produced by this mirror trajectory, but with vanishing asymptotic flux. 

\bea
x(t)&=&\frac{\kappa~\text{arcsinh}(\e^{\zeta t})}{\zeta},\\\nn
\tau (t)&=&t +\frac{ \sqrt{1-\kappa^2}\ln\left[2(1-\kappa^2)\e^{2 t \zeta}+2\sqrt{1-\kappa^{2}}\sqrt{(1+(1-\kappa^2)\e^{2 t \zeta})(1+\e^{2 t \zeta})} +2- \kappa^2 \right]}{2\zeta} \nn\\
&-&\frac{\ln\left[ 2(1+\e^{2 t \zeta}) -\kappa^2 \e^{2 t \zeta}+2\sqrt{(1+(1-\kappa^2)\e^{2 t \zeta})(1+\e^{2 t \zeta})} \right]}{2\zeta}
\eea

 The detector enters the cavity at
\begin{equation}
x_{0}=\frac{\kappa~\text{arcsinh}(1)}{\zeta} 
\end{equation}
and the setting of the parameters differs from the previous two cases. We fix $t_0$ and $\zeta$ by choosing
the initial acceleration and initial velocity to be equal to that of the Rindler case.  However  $\kappa$ is a free parameter $0<\kappa<1$ that determines the limiting final speed of the detector;  we choose it to be 0.65 for our study.  The detector takes time
  \bea
T=\frac{\ln \Big[\text{sinh}\left(\frac{L\zeta +\kappa ~\text{arcsinh}(e^{t_0 \zeta})}{\kappa}\right)\Big]}{\zeta}.
\eea
to travel the full length of the cavity, and 
  \begin{equation}
\alpha(t)=\frac{\kappa \zeta \e^{\zeta t}}{(1-(\kappa^2-1) \e^{2\zeta t})^{3/2}}
\end{equation}
is its proper acceleration.We see that this quantity vanishes at late times.

\section{Finite Flux mirror trajectories}

\begin{figure}[htp]
\centering
\includegraphics[scale=0.48]{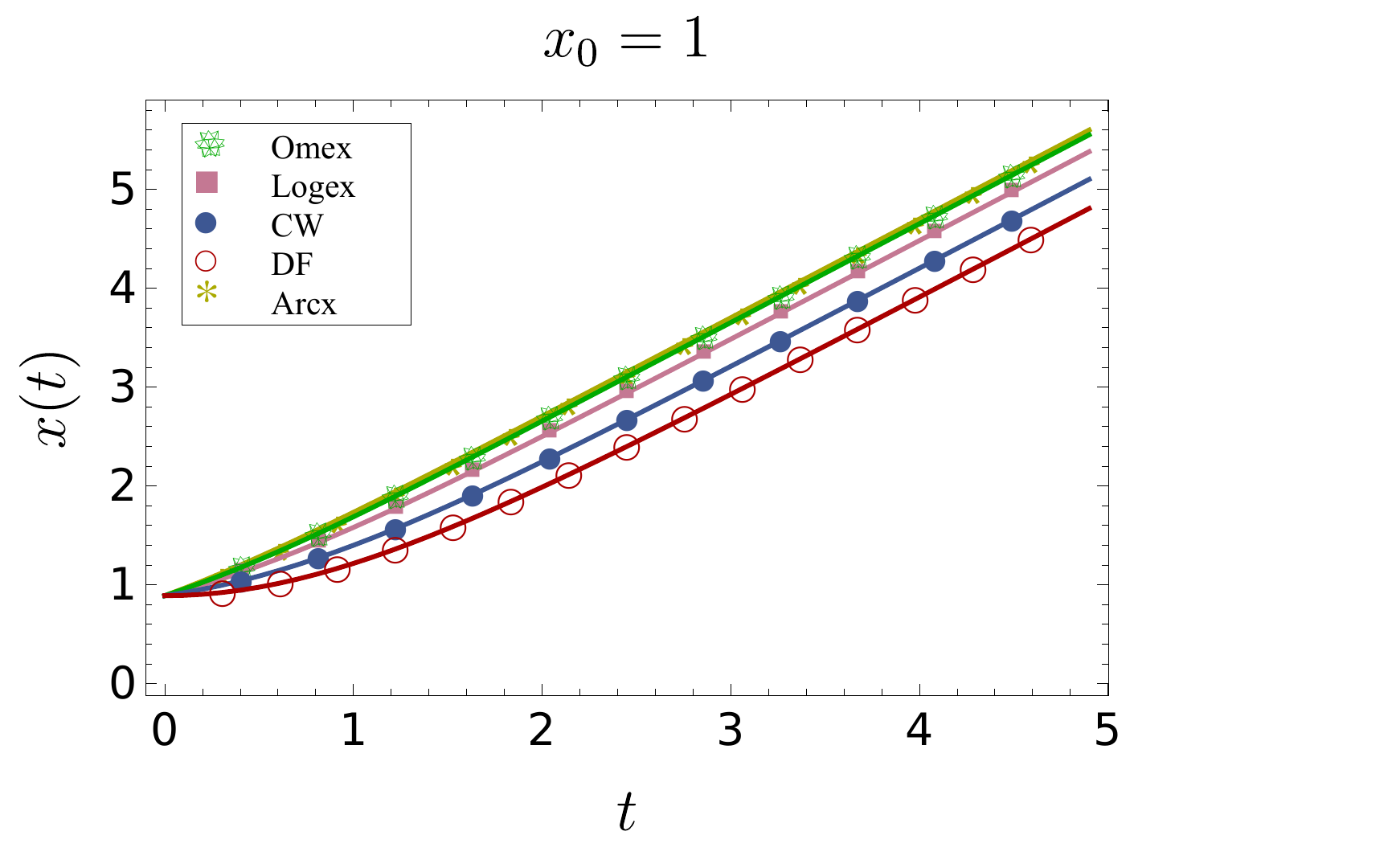}
\caption{Various trajectories of finite flux: DF, Arcx, Omex, Logex, and CW are shown. Initial position of all trajectories is chosen to be $x_0=1$ so that all trajectories and their parameters stay positive and physical. The acceleration in the Rindler case has been normalized to $\alpha_0=1$. Initial accelerations of the other trajectories are normalized to this value.}
\label{traj2-x}
\end{figure}

These trajectories have the common feature that at late times the energy flux from the mirror trajectory asymptotes to a constant value. This value can be calibrated to be equal for all such trajectories, and we shall do so here. We begin with a description of these trajectories shown in Fig.~\ref{traj2-x}. The velocities of these trajectories are illustrated in Fig.~1(b) in the main text.

\subsection{Davies-Fulling}
 
This is one of the earliest mirror trajectories studied, and was used to demonstrate that a Planck spectrum from a moving mirror can be obtained that is analogous to that found for black hole evaporation. The Davies-Fulling mirror trajectory has a time-dependent acceleration that is asymptotically null. The parametrization of this trajectory is given by \cite{Goodthesis} 
\begin{align}
&x(t)=\frac{\ln  [\textrm{cosh}(\xi t)] }{\xi},\quad 
\tau(t) = \frac{2 \textrm{arctan}(\textrm{tanh}(\frac{\xi t}{2}))}{\xi}\quad 
\end{align}
where $\xi$ is a positive constant whose relationship to the proper acceleration   is 
\begin{equation}
\alpha(t)=\frac{\xi}{\sqrt{\textrm{sech}(\xi t)}}.
\end{equation}
Setting  the initial acceleration and   velocity of the detector to be that of the corresponding Rindler detector,
we fix both $\xi$ and $t_0$, obtaining 
  \be
T=\frac{\textrm{arccosh}\left(\frac{1+\e^{2 \xi t0}}{2\e^{\xi (t_0-L)}}\right)}{\xi},
\ee
for time this detector spends  in the cavity.

\subsection{Carlitz-Willey}

This trajectory, the CW trajectory, is of physical interest insofar as it simulates an eternal black hole that  evaporates thermally  at  fixed temperature.  The mirror trajectory has constant  energy flux (and thus a divergent amount of total energy).  There is a thermal spectrum at all times and the  Bogoliubov coefficients can be computed exactly and analytically.   It does not make use of any late time approximations.

The CW trajectory is parametrized as
\begin{align}
&x(t) = t+\frac{W(\e^{-2kt})}{k},  \quad
\tau(t) = -2\frac{\sqrt{W(\e^{-2kt})}}{k}
\end{align}
and as $t\to -\infty$ is asymptotically null with zero proper acceleration \cite{Goodthesis}. The quantities
$k$ and $t_0$ are fixed as before by  requiring equality of the initial acceleration and velocity with the Rindler case, and 
\bea
T&=&(L+t_0)\\\nn
&+&\frac{W(-\e^{-2k(L-t_0)}W(\e^{-2 k t_0})^2)+W(\e^{-2 k t_0})}{k},
\eea
is time that it takes for the detector to traverse the cavity  
and 
 \begin{equation}
\alpha(t)=\frac{k}{2\sqrt{W(\e^{-2kt})}},
\end{equation}
is its proper acceleration.

\subsection{Arcx}

This trajectory is analogous to the Davies-Fulling trajectory, with the advantage that it has a static start but with velocity and acceleration continuous at all times, allowing for a solution that is valid globally \cite{Goodthesis}.
The mirror trajectory has thermal late time emission and infinite acceleration, and an energy flux that also asymptotes to a constant value in the distant future. It is given by
\begin{align}
&x(t) = \frac{\text{arcsinh}(\e^{kt})}{k}, \quad
\tau(t) = -\frac{\text{arctanh}\left(\frac{1}{\sqrt{1+\e^{2kt}}}\right)}{k}
\end{align}
where 
 \begin{equation}
\alpha(t) =k~\e^{kt},
\end{equation}
and
\bea
T=\frac{\log \big[\text{sinh}\big(kL+\text{arcsinh}(\e^{k t})\big)\big]}{k}
\eea
are the respective proper acceleration and time spent by the detector in the cavity, with $k$ and  $t_0$
calibrated to the Rindler case as before.

\subsection{Logex}

Unlike the other mirror trajectories, Logex emits a pulse of energy flux before asymptoting to the CW value
\cite{Goodthesis}.  This trajectory that starts off asymptotically static is always accelerating and is
given by 
\bea
x(t)&=&\frac{\ln(1+\e^{2kt})}{2k},\\\nn
\!\!\!\!\!\!\!\!\!\tau(t)&=&\frac{1}{2k}\!\left(\!2 {\arctan}\left(\!\sqrt{1+2\e^{2kt}}\right) + \ln\!\left[\frac{\sqrt{1+2\e^{2kt}}-1}{\sqrt{1+2\e^{2kt}}+1} \right]\right)
\eea
where 
\be
T=\frac{\ln(\e^{2k(L+t_0)}+\e^{2kL}-1)}{2k}
\ee
is the time spent in the cavity and 
\bea
\alpha(t)=\frac{2 k \e^{2kt}(1+\e^{2kt})}{(1+2\e^{2kt})^{3/2}}
\eea
is the  proper acceleration, which  diverges at late times.  As before, calibration with the Rindler trajectory fixes
 $k$ and $t_0$.   

\subsection{Omex}

The last trajectory we study is the Omex mirror trajectory.  This one is of considerable interest since its Bogoliubov coefficients are identical to those of a Schwarzschild black hole truncated to two spacetime dimensions \cite{Goodthesis}. Its energy flux asymptotes to a constant value in the distant future.

The Omex  trajectory is of similar form to that of the Carlitz-Willey trajectory, but is  asymptotically static in the distant past, and is given by
\bea
x(t)&=&t+\frac{W(\e^{-2kt})}{2k},\\\nn
\tau(t)&=&-\frac{1}{2k}\Bigg( \sqrt{(2+W(\e^{-2kt}))W(\e^{-2kt})}\\\nn
\!\!\!\!&+&\ln\left[1+W(\e^{-2kt})+ \sqrt{(2+W(\e^{-2kt}))W(\e^{-2kt})}\right]\!\Bigg),
\eea
where $k$ and $t_0$ are determined from calibration with the Rindler trajectory as before.    The detector spends a time
\bea
T=\frac{2k(L+t_0)+W(\e^{-2 k t_0})(1-\e^{-2kL})}{2 k},
\eea
travelling the proper length  $L$ of the cavity and 
\begin{equation}
\alpha(t)=\frac{2k}{\sqrt{W(\e^{-2kt})(2+W(\e^{-2kt}))^{3}}}
\end{equation} 
is its proper  acceleration.

\section{Detector responses: Vanishing and Non-vanishing responses}

\begin{figure*}[h]
\centering
\includegraphics[scale=0.48]{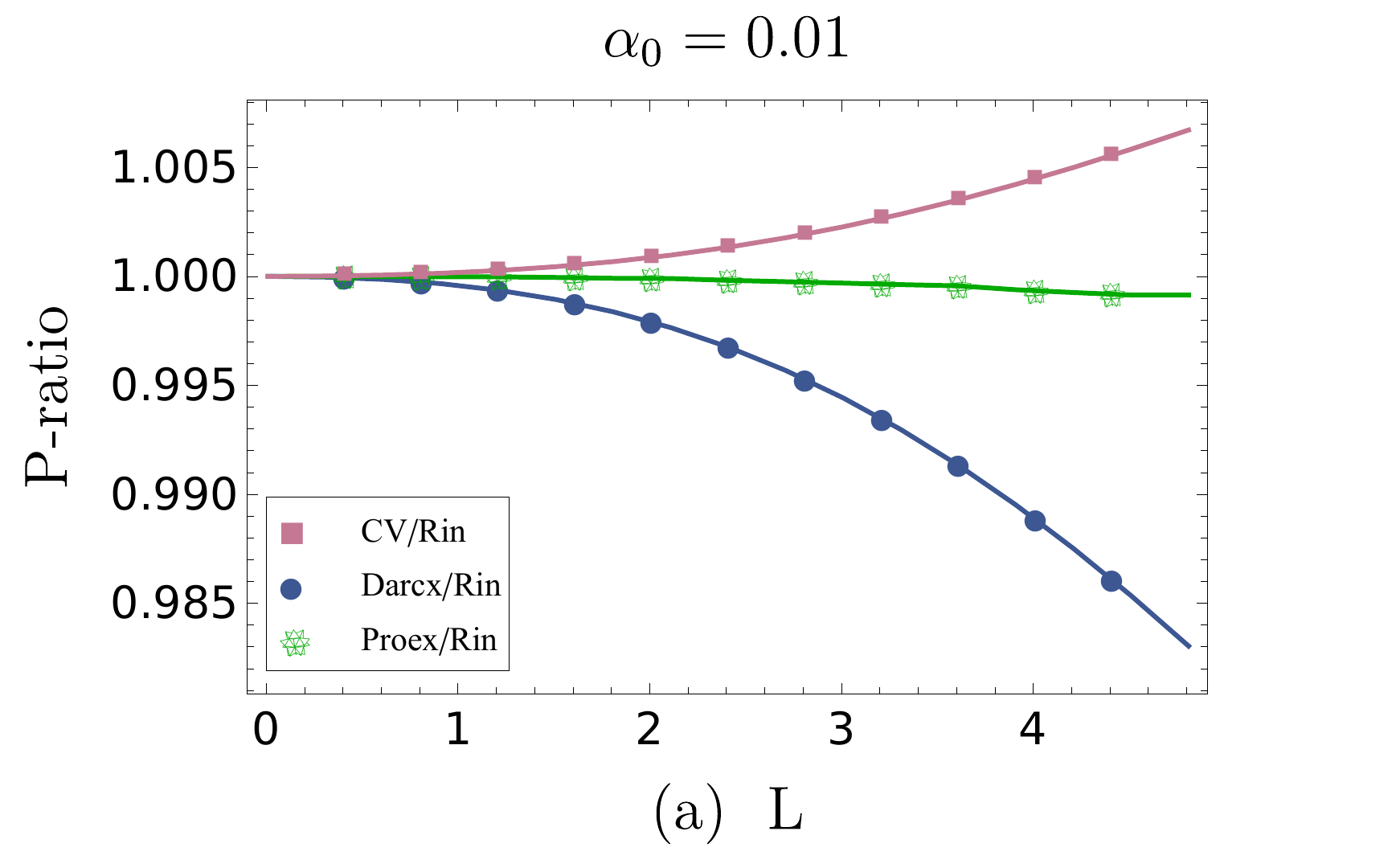}
\quad
\includegraphics[scale=0.48]{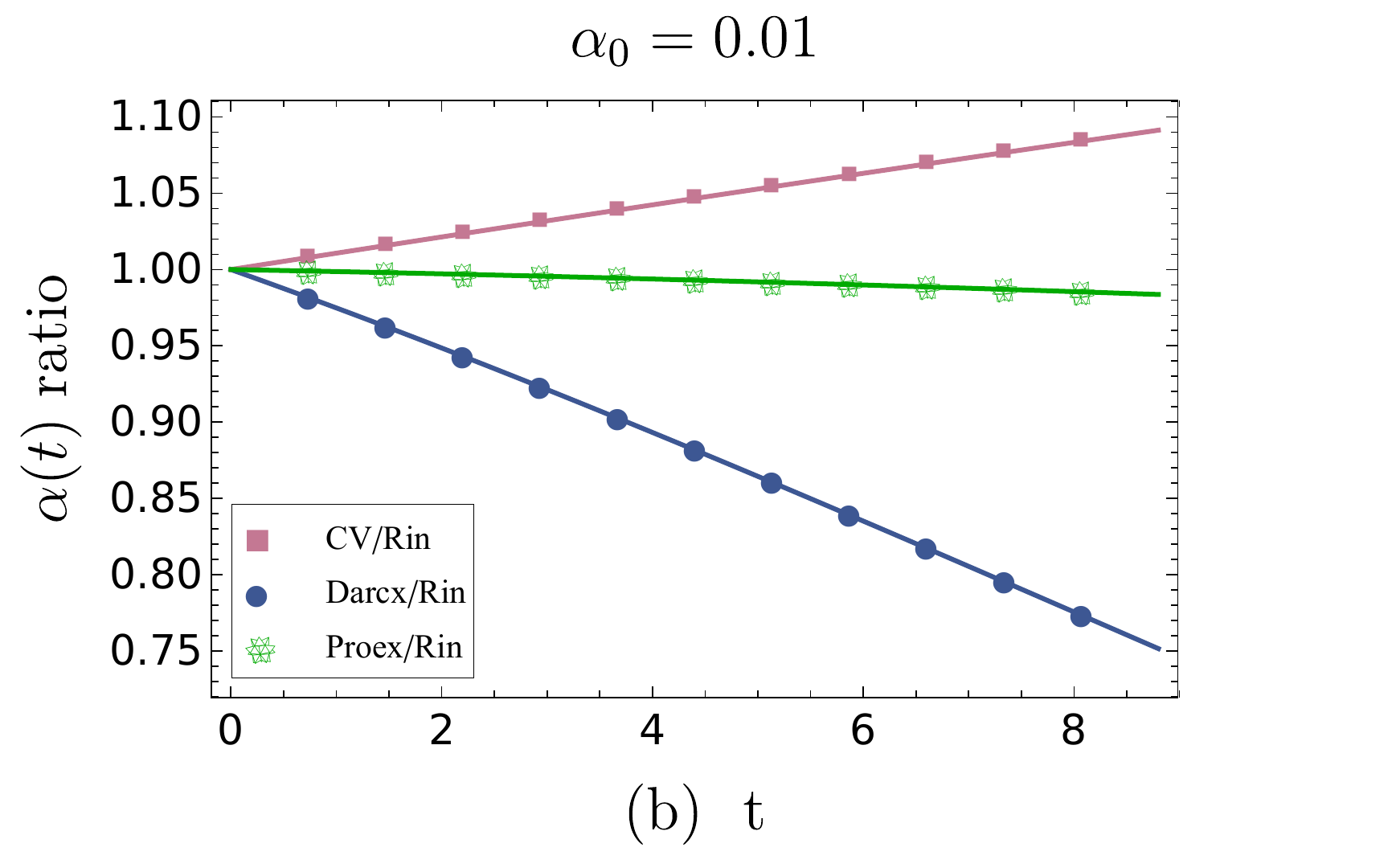}
\quad
\includegraphics[scale=0.48]{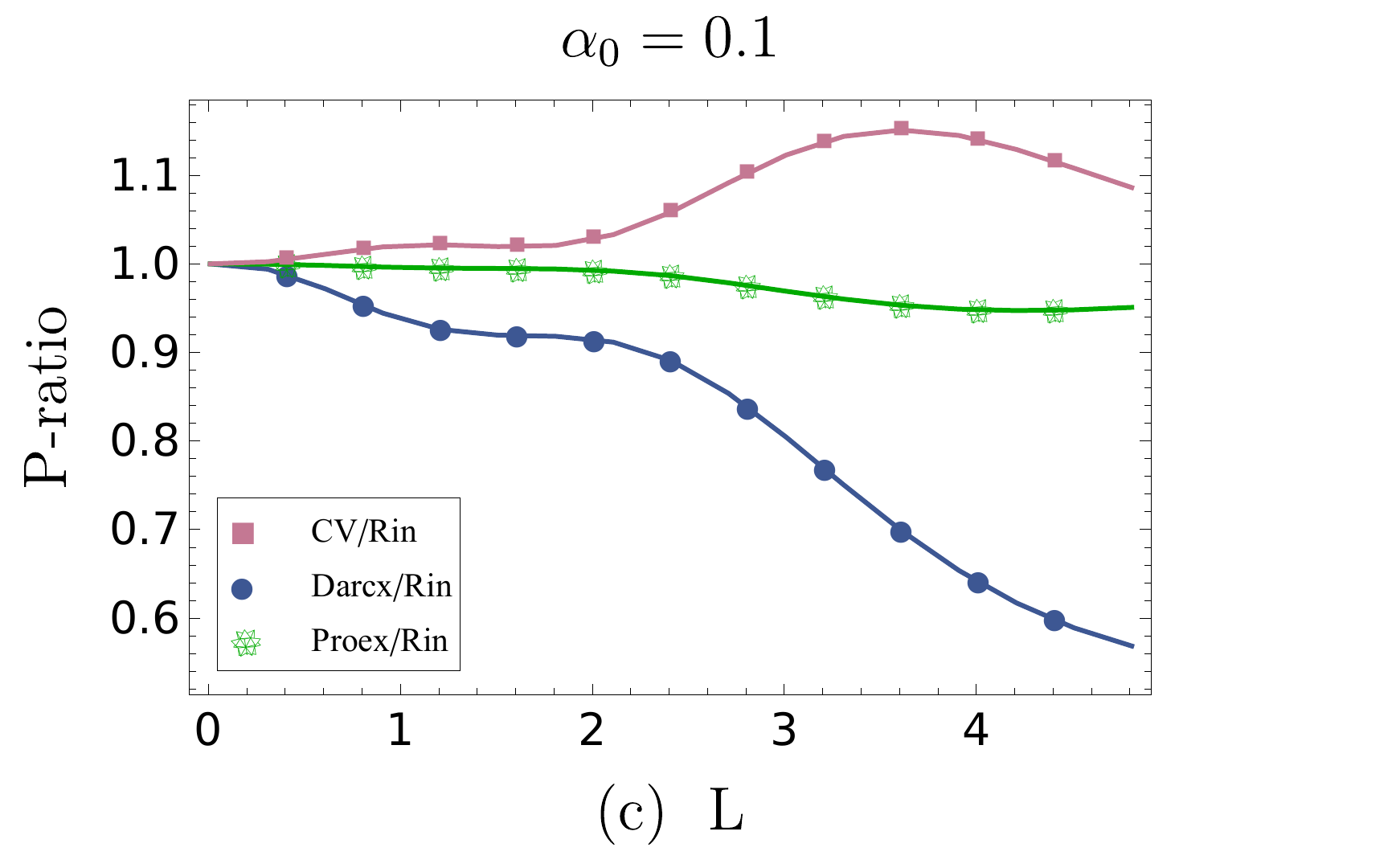}
\quad
\includegraphics[scale=0.48]{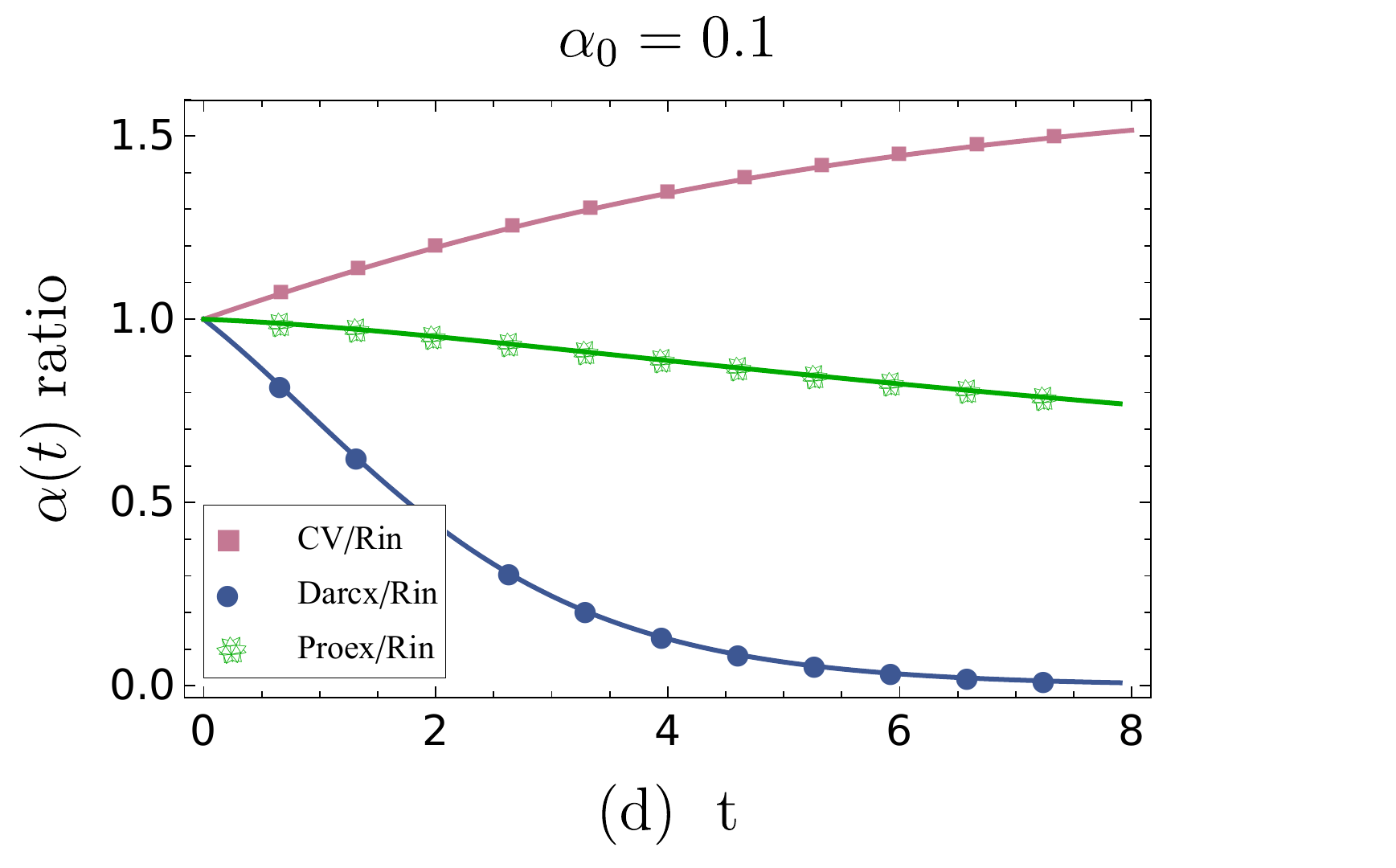}
\quad
\includegraphics[scale=0.48]{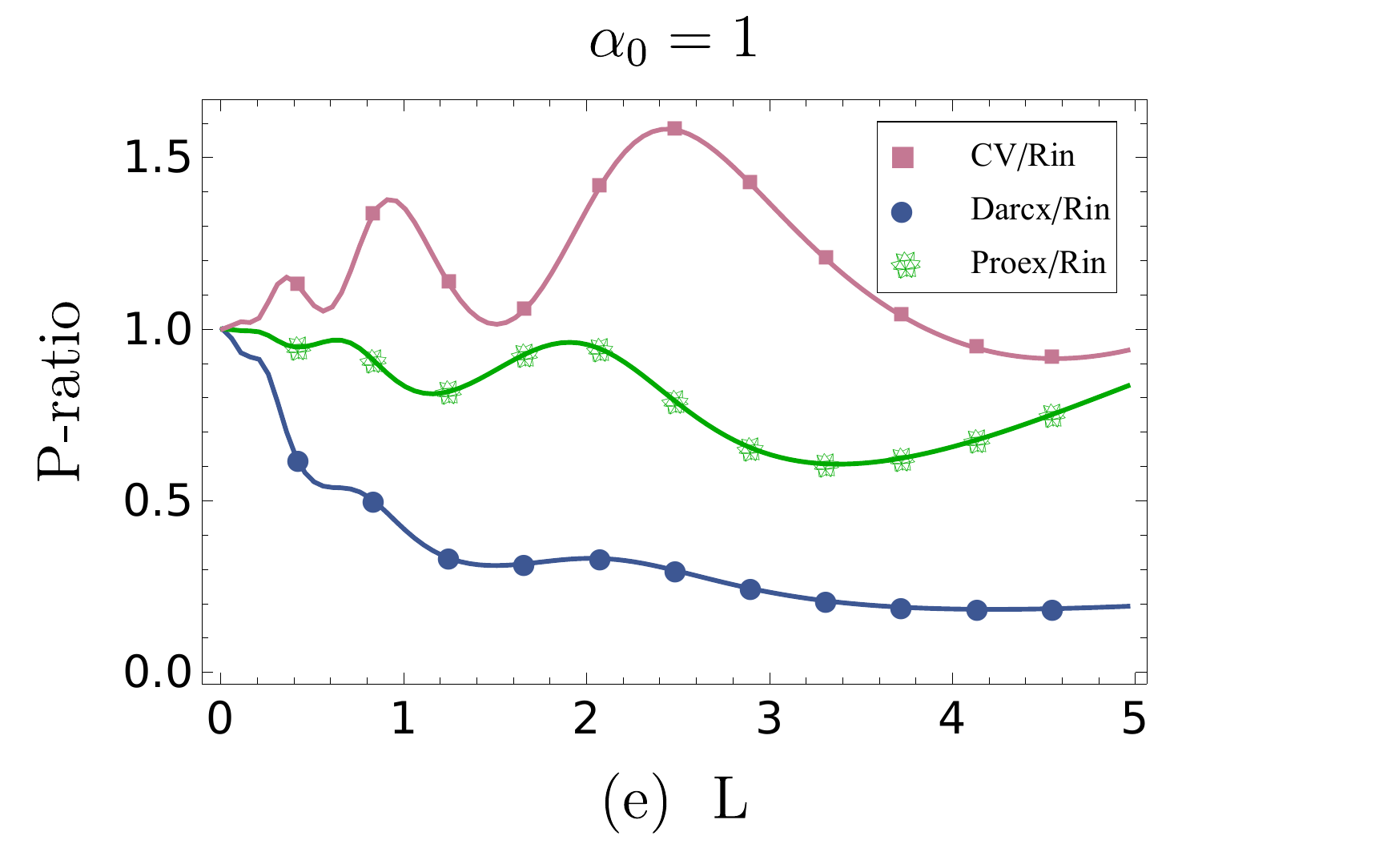}
\quad
\includegraphics[scale=0.48]{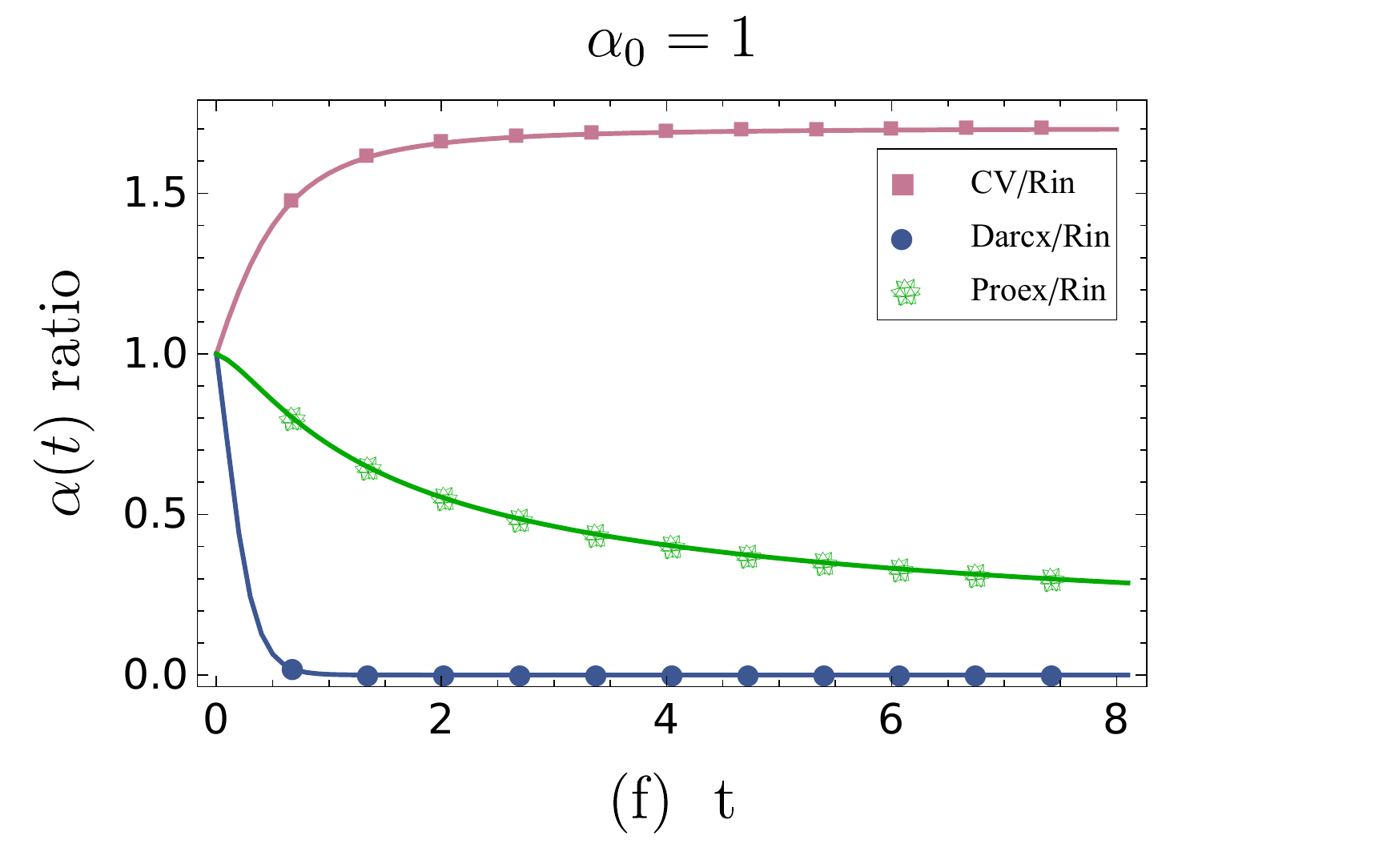}
\caption{A comparison of the excitation probability (left column)  and proper acceleration (right column) of detectors traveling on non-uniformly accelerated trajectories with vanishing flux with the excitation probability of a uniformly accelerated detector. The three ratios are Costa-Villalba to Rindler (pink square), Darcx to Rindler (blue circle), and Proex to Rindler (green star). Each plot represents the behaviour of these ratios as a function of the length of the cavity that they are traveling through for different initial accelerations: a) $\alpha_0=0.01$, c) $\alpha_0=0.1$, and e) $\alpha_0=1$.}
\label{vanish}
\end{figure*}

\begin{figure*}[htp]
\centering
\includegraphics[scale=0.48]{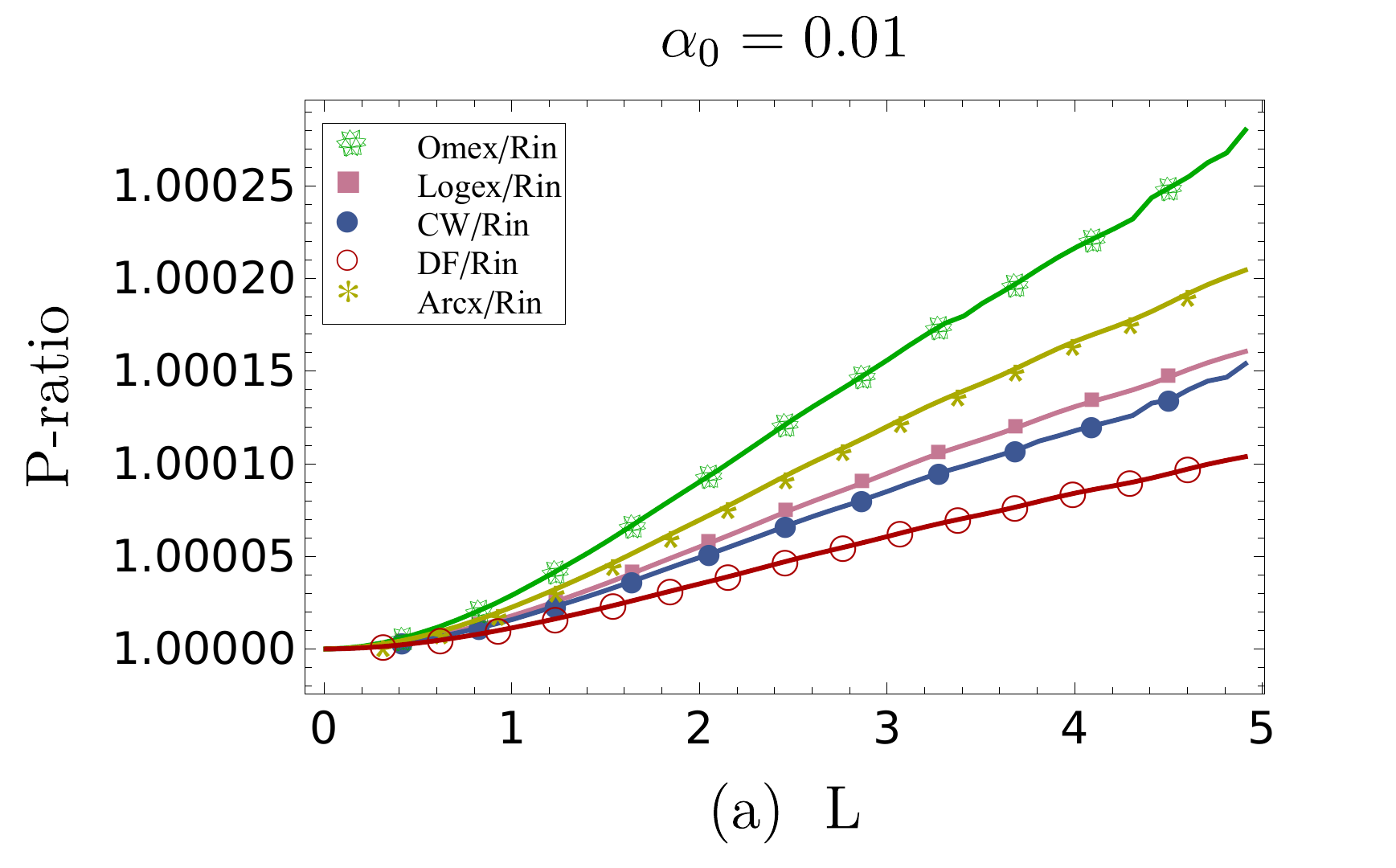}
\quad
\includegraphics[scale=0.48]{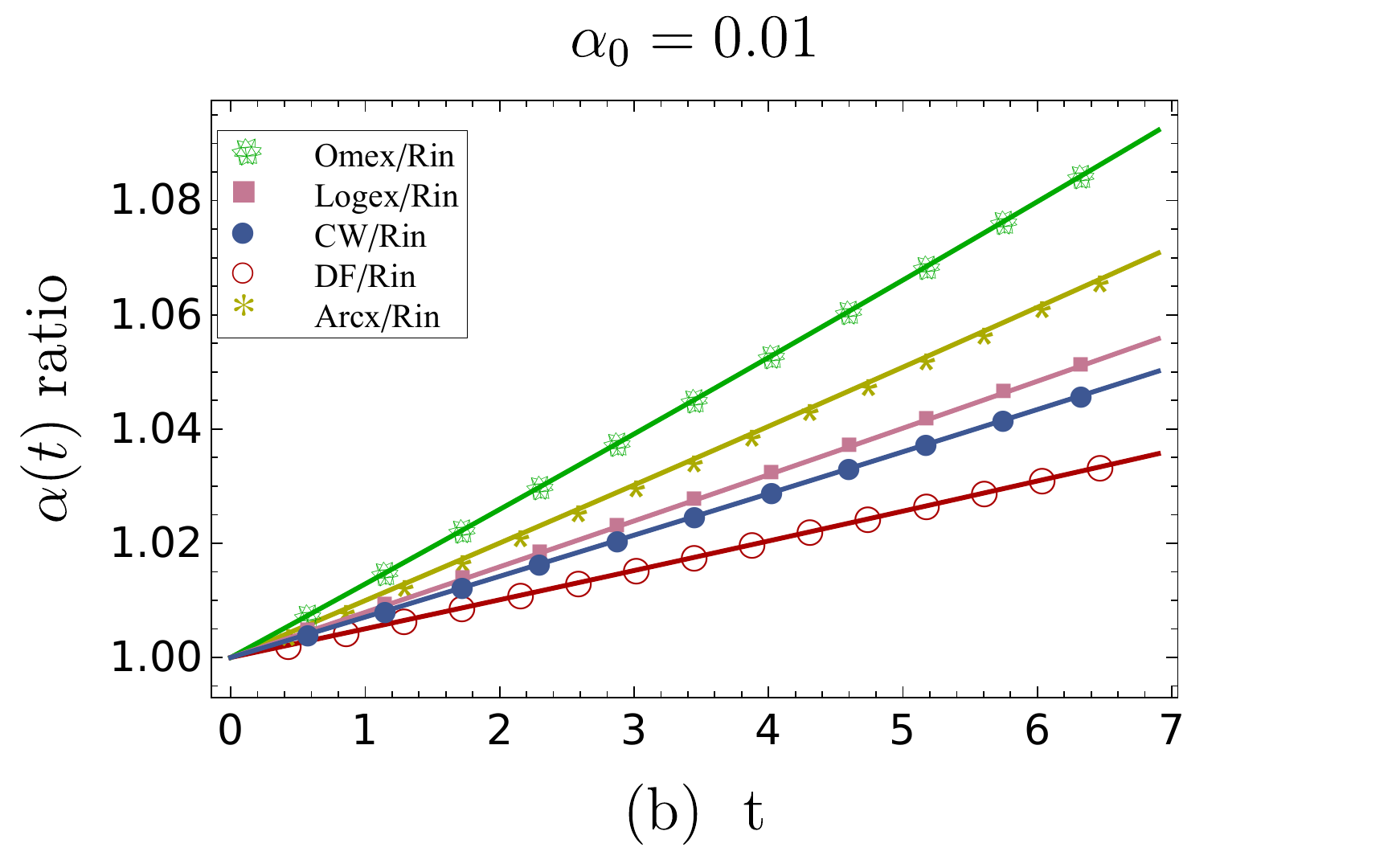}
\quad
\includegraphics[scale=0.48]{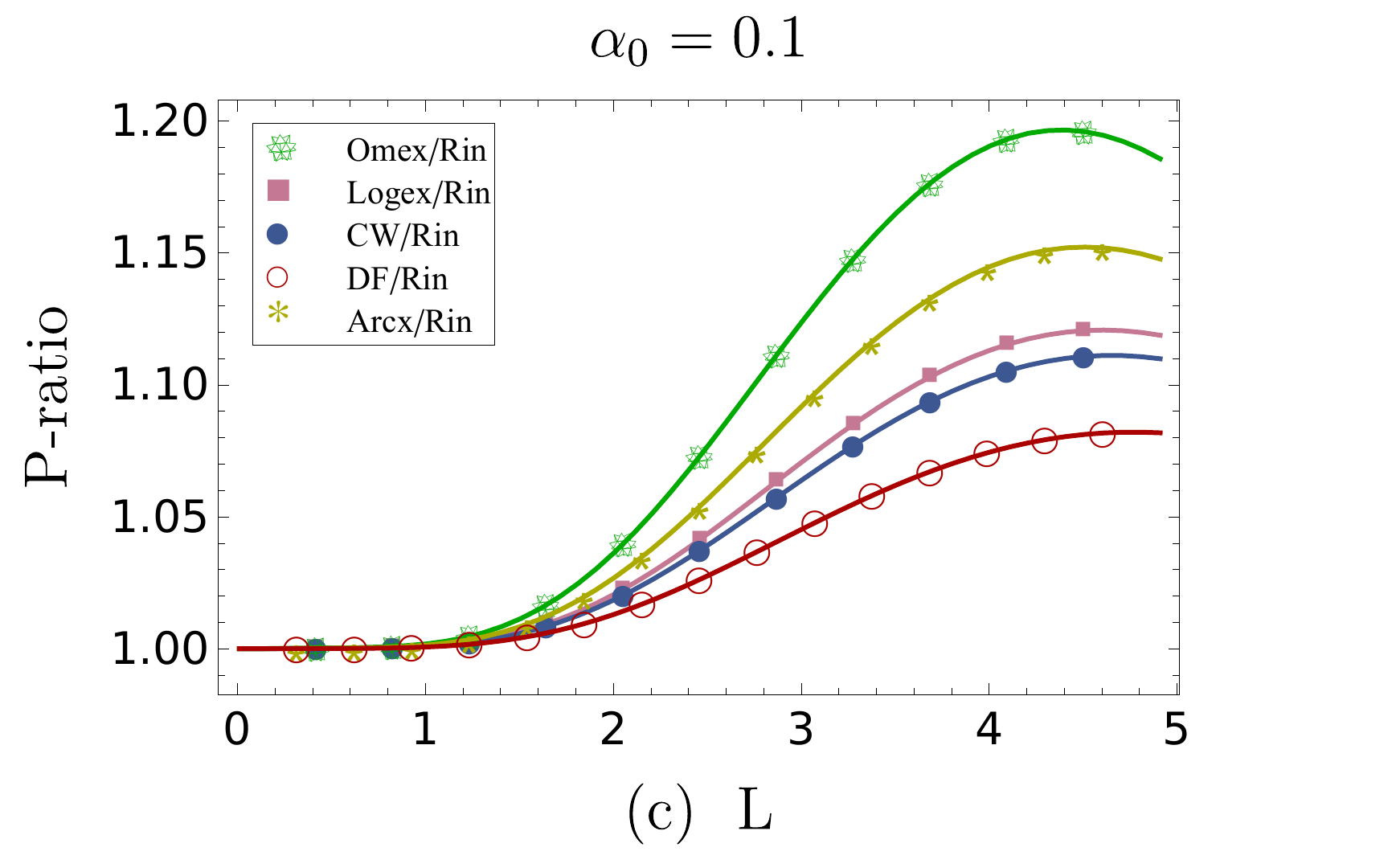}
\quad
\includegraphics[scale=0.48]{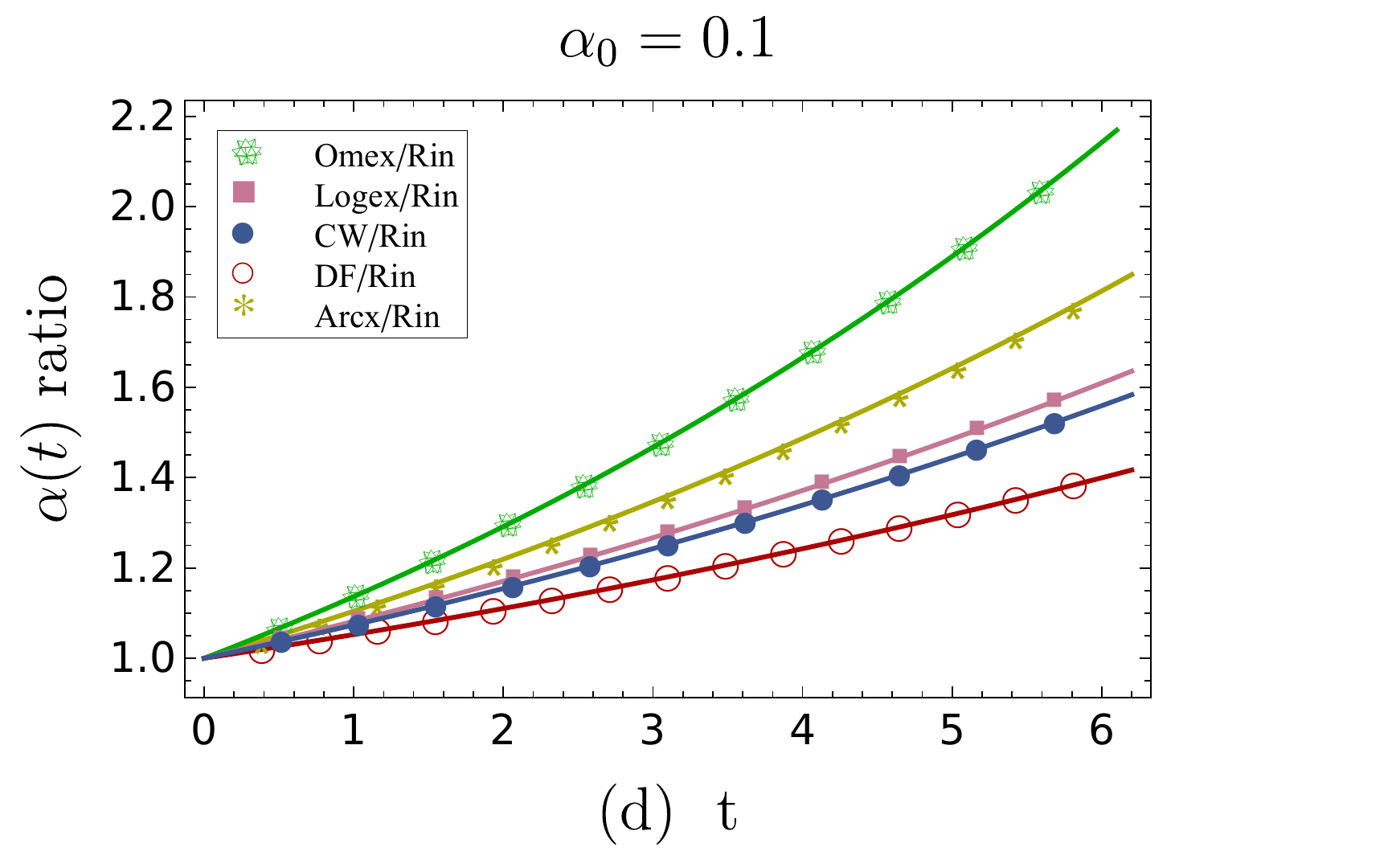}
\quad
\includegraphics[scale=0.48]{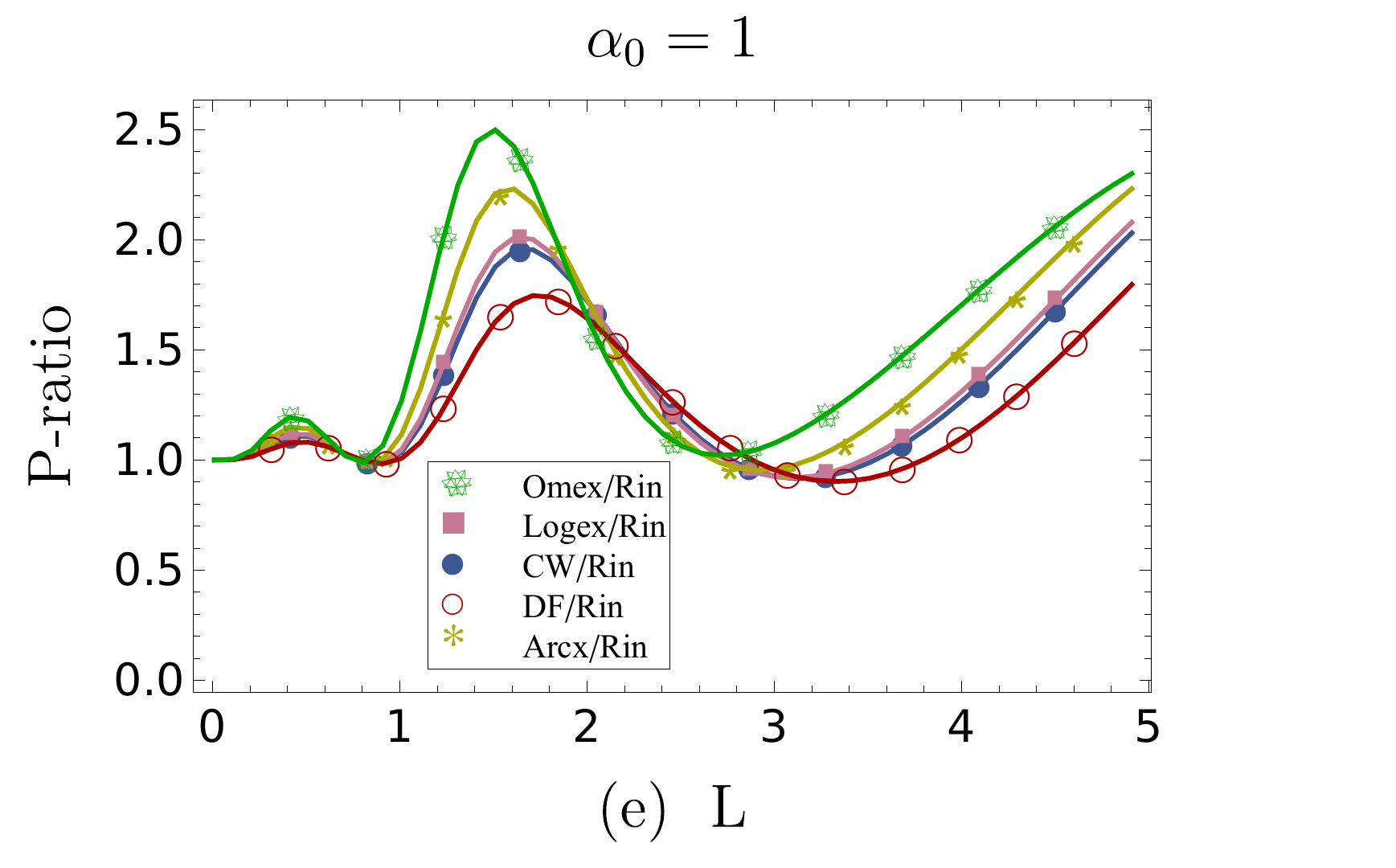}
\quad
\includegraphics[scale=0.48]{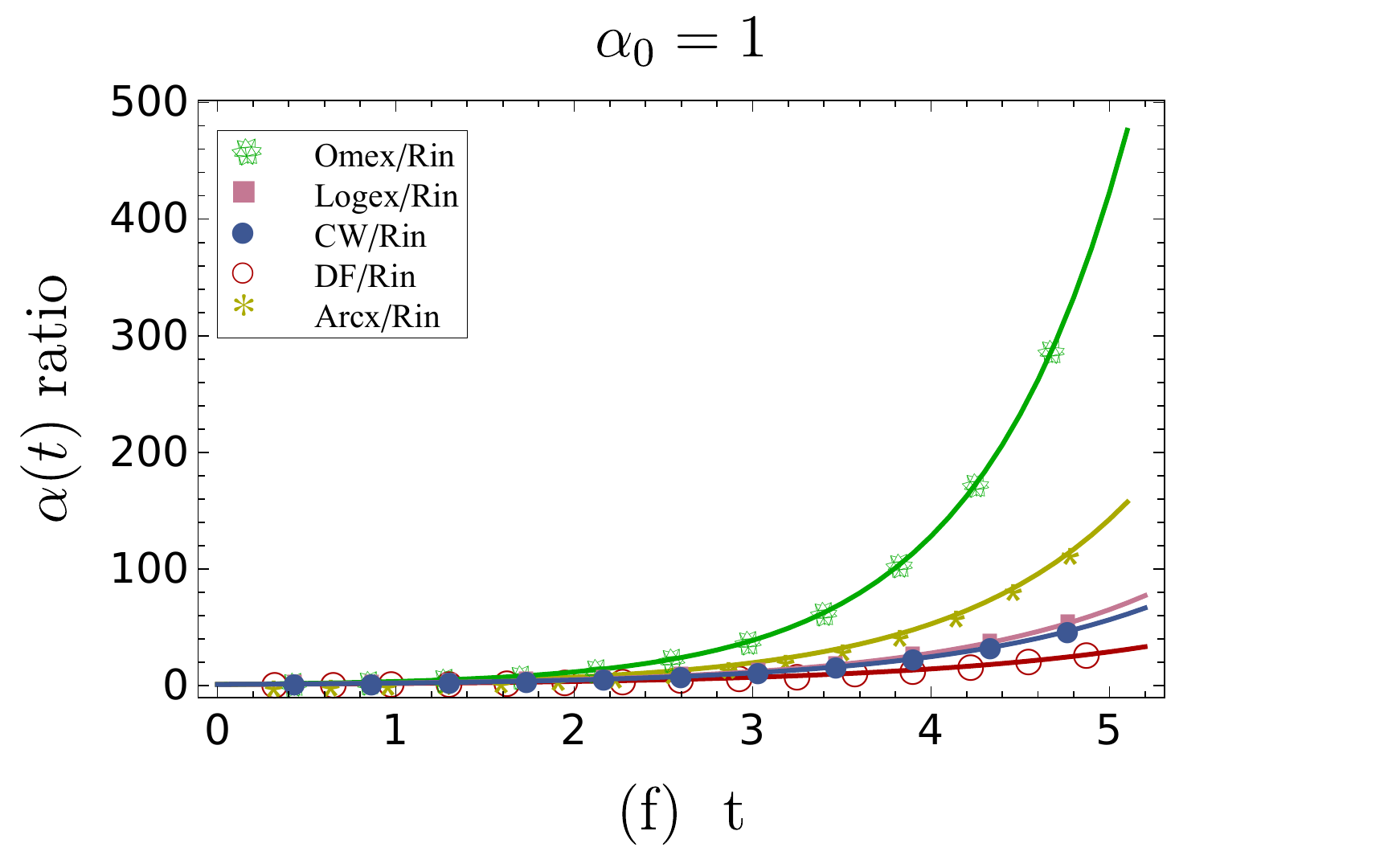}
\caption{Comparing the excitation probability of detectors traveling on non-uniformly accelerated trajectories with non-vanishing flux with the excitation probability of a uniformly accelerated detector. The five ratios are Omex to Rindler (in green flake), Logex to Rindler (in pink square), CW to Rindler (in blue circle), DF to Rindler (in red circle), and Arcx to Rindler (in yellow star). Each plot represents the behaviour of these ratios as a function of the length of the cavity that they are traveling through for different initial accelerations: a) $\alpha_0=0.01$, c) $\alpha_0=0.1$, and e) $\alpha_0=1$.}
\label{nonvanish}
\end{figure*}

In Figs.~\ref{vanish} and~\ref{nonvanish} we plot the responses of detectors traveling along each of the trajectories listed above for increasing acceleration parameters, alongside plots showing how the acceleration increases as a function of time.  The Omex trajectory provides the greatest contrast with the Rindler case.

\newpage
\twocolumngrid
\bibliography{cavity_refs}
\bibliographystyle{apsrev4-1}

\end{document}